\documentclass[acmsmall,screen,nonacm]{acmart}

\usepackage{amsmath}
\newcommand{\AlgName}[1]{\ensuremath{\text{{\sf #1}}}}

\usepackage{appendix}
\usepackage{afterpage}
\usepackage{float}
\usepackage{placeins}
\usepackage{graphicx}
\usepackage{balance}  

\usepackage{caption}
\usepackage{subcaption}

\usepackage[utf8]{inputenc}
\usepackage{url}
\usepackage{algorithm}
\usepackage{algorithmic}
\usepackage{numprint}
\npdecimalsign{.} 

\usepackage{hyperref}
\usepackage{xspace}
\usepackage{wrapfig}
\usepackage{graphics}
\usepackage{todonotes}
\usepackage{booktabs}
\usepackage{microtype}
\definecolor {infocolor} {rgb} {0.6,0.6,0.6}

\usepackage{rotating}
\usepackage{lscape} 
\usepackage{threeparttable}
\usepackage{hvfloat}

\usepackage{multirow}

\usepackage{doi}

\usepackage{tikz}
\usetikzlibrary{shapes.arrows,chains}
\usetikzlibrary{decorations.pathreplacing}

\usepackage{framed,enumitem}

\def\MdR{\ensuremath{\mathbb{R}}}

\newcommand{\Id}[1]{\texttt{\detokenize{#1}}}

\newcommand{\ie}{i.\,e.,\xspace}

\newcommand{\etal}{et~al.\xspace}

\usepackage{color}

\def\comment#1{}

\def\withcomments{
	\newcounter{mycommentcounter}
	\def\comment##1{\refstepcounter{mycommentcounter}%
		\ifhmode%
		\unskip%
		{\dimen1=\baselineskip \divide\dimen1 by 2 %
			\raise\dimen1\llap{\tiny\bfseries \textcolor{red}{-\themycommentcounter-}}}\fi%
		\marginpar[{\renewcommand{\baselinestretch}{0.8}%
			\hspace*{3em}\begin{minipage}{5em}\footnotesize [\themycommentcounter]: \raggedright ##1\end{minipage}}]{\renewcommand{\baselinestretch}{0.8}%
			\begin{minipage}{5em}\footnotesize [\themycommentcounter]: \raggedright ##1\end{minipage}}}
}

\newcommand{\Xcomment}[1]{}

\usepackage{amsfonts}

\newcommand{\set}[1]{\left\{ #1\right\}}
\newcommand{\gilt}{:}
\newcommand{\sodass}{\,:\,}
\newcommand{\setGilt}[2]{\left\{ #1\sodass #2\right\}}

\newcommand{\realrange}[2]{\left[#1, #2\right]}

\newcommand{\unitrange}[2]{\realrange{0}{1}}

\newcommand{\llabel}[1]{\label{\labelprefix:#1}}
\newcommand{\labelprefix}{} 

\newcommand{\discussionsize}{\small}

\newenvironment{code}{\noindent
	\begin{tabbing}%
		\hspace{2em}\=\hspace{2em}\=\hspace{2em}\=\hspace{2em}\=\hspace{2em}\=%
		\hspace{2em}\=\hspace{2em}\=\hspace{2em}\=\hspace{2em}\=\hspace{2em}\=%
		\kill}{\end{tabbing}}

\newcommand{\labelcommand}{}
\newcommand{\captiontext}{}
\newsavebox{\codeparam}
\newcounter{lineNumber}
\newenvironment{disscodepos}[3]{%
	\renewcommand{\labelcommand}{#2}%
	\renewcommand{\captiontext}{#3}%
	\sbox{\codeparam}{\parbox{\textwidth}{#3}}%
	\begin{figure}[#1]\begin{center}\begin{code}\setcounter{lineNumber}{1}}{%
		\end{code}\end{center}\caption{\llabel{\labelcommand}\captiontext}\end{figure}}

	{\end{disscodepos}}

\newcommand{\Is}       {:=}

\newdimen\endofsize\endofsize=0.5em

\newcommand{\mytitle}{Buffered Streaming Graph Partitioning}

\AtBeginDocument{%
  \providecommand\BibTeX{{%
    \normalfont B\kern-0.5em{\scshape i\kern-0.25em b}\kern-0.8em\TeX}}}

\begin{document}

\title{\mytitle}

\author{Marcelo Fonseca Faraj}
\email{marcelofaraj@informatik.uni-heidelberg.de}
\orcid{0000-0001-7100-236X}
\affiliation{%
  \institution{Heidelberg University}
  \city{Heidelberg}
  \country{Germany}
}

\author{Christian Schulz}
\email{christian.schulz@informatik.uni-heidelberg.de}
\orcid{0000-0002-2823-3506}
\affiliation{%
  \institution{Heidelberg University}
  \city{Heidelberg}
  \country{Germany}
}

%
%
%
%
%
%

\renewcommand{\shortauthors}{Faraj and Schulz}

\begin{abstract}
  Partitioning graphs into blocks of roughly equal size is a widely used tool when processing large graphs. Currently there is a gap observed in the space of available partitioning algorithms. On the one hand, there are streaming algorithms that have been adopted to partition massive graph data on small machines. In the streaming model, vertices arrive one at a time including their neighborhood and then have to be assigned directly to a block. These algorithms can partition huge graphs quickly with little memory, but they produce partitions with low solution quality. On the other hand, there are offline (shared-memory) multilevel algorithms that produce partitions with high quality but also need a machine with enough memory to partition~huge~networks. In this work, we make a first step to close this gap by presenting an algorithm that computes significantly improved partitions of huge graphs using a single machine with little memory in streaming setting. First, we adopt the buffered streaming model which is a more reasonable approach in practice. In this model, a processing element can store a buffer, or batch, of nodes (including their neighborhood) before making assignment decisions. When our algorithm receives a batch of nodes, we build a model graph that represents the nodes of the batch and the already present partition structure. This model enables us to apply multilevel algorithms and in turn compute much higher quality solutions of huge graphs on cheap machines than previously possible. To partition the model, we develop a multilevel algorithm that optimizes an objective function that has previously shown to be effective for the streaming setting. Surprisingly, this also removes the dependency on the number of blocks $k$ from the running time compared to the previous state-of-the-art. Overall, our algorithm computes, on average, $75.9\%$ better solutions than \AlgName{Fennel} using a very small buffer size. In addition, for large values of $k$ our algorithm becomes faster than \AlgName{Fennel}. 
\end{abstract}





\maketitle

\section{Introduction}
\label{sec:introduction}

Complex graphs are increasingly being used to model phenomena such as social networks, data dependency in applications, citations of papers, and biological systems like the human brain. 
Often these graphs are composed of billions of entities that give rise to specific properties and structures.
As a concrete example to cope with such graphs, graph databases~\cite{DBLP:journals/vldb/DemirciFA19} and graph processing frameworks  \cite{ching2015one, malewicz2010pregel, gonzalez2012powergraph} can be used to store a graph, query it, and provide other operations.
If the graphs become too large, then they have to be distributed over many machines in order for the system to provide scalable operations.
A key operation for scalable computations on huge graphs is to partition its components among $k$ processing elements (PEs) such that each PE receives roughly the same amount of components and the communication between PEs in the underlying application is minimized.
In the distributed setup, each PE operates on some portion of the graph and communicates with other PEs through message-passing.
This operation is naturally modeled by the graph partitioning problem, which computes a partition of the graph into $k$ blocks such that the blocks have roughly the same size and the number of edges crossing blocks, \ie communication, is minimized. 
Graph partitioning is NP-complete \cite{Garey1974} and there can be no approximation algorithm with a constant ratio factor for general graphs~\cite{BuiJ92}. 
Thus, heuristic algorithms are used in practice.  

There has been an extensive body of work in the area of graph partitioning. 
Roughly speaking, there are streaming algorithms, internal memory (shared-memory parallel) algorithms and distributed memory parallel algorithms.
However, currently there is a gap observed in the design space of available partitioning algorithms.
First of all, the most popular streaming approach in literature is the one-pass model, in which vertices arrive one at a time including their neighborhood and then have to be permanently assigned to blocks. 
Algorithms based on this model can partition huge graphs quickly with little memory, but compute rather low-quality partitions.
To improve partition quality, the graph can be restreamed while the one-pass algorithm updates block assignment, but this is still far behind offline approaches. 
Offline multilevel algorithms such as \AlgName{KaHIP}~\cite{kaffpa} or \AlgName{Metis}~\cite{karypis1998fast} are widely known and can produce partitions with high quality.
Nevertheless they cannot partition huge graphs unless a machine with sufficient memory is used and hence can not be used for example for preprocessing in out-of-core algorithms.
Lastly, distributed algorithms are able to partition huge graphs successfully and compute high-quality solutions.
However, they require a large amount of computational resources and typically access to a supercomputer, which can be infeasible depending on the application.
Moreover, a distributed partitioning algorithm needs to split the input graph among different machines before actually partitioning it.
When the graph is too large, this preliminary partition has to be generated on the fly by a stream partitioning algorithm while loading the graph.
Other natural applications for stream partitioning include distributed graph processing systems based on a load-compute-store logic such as MapReduce~\cite{dean2008mapreduce} and Giraph~\cite{ching2015one}, and systems which support native graph-as-a-stream computations such as Kineograph~\cite{cheng2012kineograph}, and Apache~Flink~\cite{carbone2015apache}.

\emph{Contribution.} In this work, we start to fill the gap currently observed for the existing graph partitioning algorithms.
We propose an algorithm that can compute significantly better partitions of huge graphs than the currently available streaming algorithms while using a single machine without a lot of memory.
We adopt the buffered streaming model which allows a buffer of nodes to be received and stored before making assignment decisions.
Our algorithm is carefully engineered to produce partitions of improved quality by using a sophisticated multilevel scheme on a compressed model of the buffer \emph{and} the already assigned vertices. 
Our multilevel algorithm optimizes for the same objective as the previous state-of-the-art \AlgName{Fennel}. 
However, due to the multilevel scheme used on the compressed model, our local search algorithms have a global view on the optimization problem and hence compute better solutions overall. 
Lastly, using the multilevel scheme reduces the time complexity from $O(nk+m)$ of \AlgName{Fennel} to $O(n+m)$, where $k$ is the number of blocks a graph has to be partitioned in.
To this end, experiments indicate that our algorithm can partition huge networks on machines with small memory while computing \emph{better} solutions than the previous state-of-the-art in the streaming setting. 
At the same time our algorithm is \emph{faster} than the previous state-of-the-art for larger values of blocks~$k$.

\begin{figure}[t]
	\centering
	\includegraphics[width=0.6\textwidth]{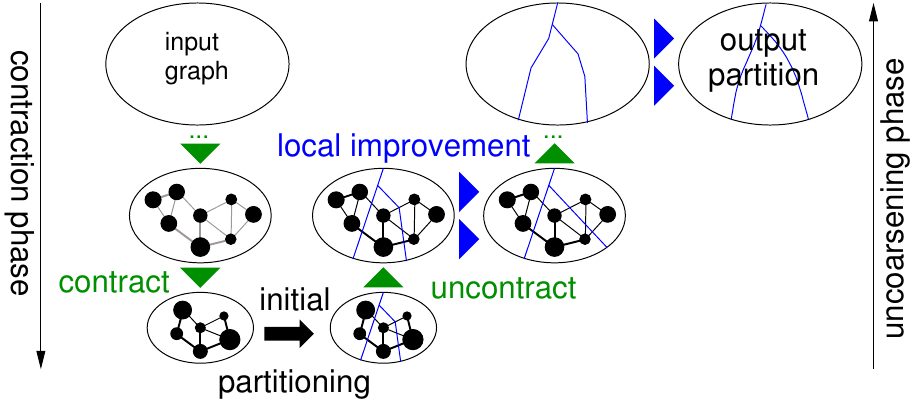}
	\caption{Multilevel graph partitioning. The graph is recursively contracted to achieve smaller graphs. After the coarsest graph is initially partitioned, a local search method is used on each level to improve the partitioning induced by the coarser level.}
	\label{fig:mgp}
        \vspace*{-1cm}
\end{figure}

\section{Preliminaries}
\label{sec:preliminaries}

\subsection{Basic Concepts}
\label{subsec:basic_concepts}

Let $G=(V=\{0,\ldots, n-1\},E)$ be an \emph{undirected graph} with no multiple or self edges allowed, such that $n = |V|$ and $m = |E|$.
Let $c: V \to \MdR_{\geq 0}$ be a node-weight function, and let $\omega: E \to \MdR_{>0}$ be an edge-weight function.
We generalize $c$ and $\omega$ functions to sets, such that $c(V') = \sum_{v\in V'}c(v)$ and $\omega(E') = \sum_{e\in E'}\omega(e)$.
Let $N(v) = \setGilt{u}{\set{v,u}\in E}$ denote the neighbors of $v$.
A graph $S=(V', E')$ is said to be a \emph{subgraph} of $G=(V, E)$ if $V' \subseteq V$ and $E' \subseteq E \cap (V' \times V')$. 
When $E' = E \cap (V' \times V')$, $S$ is an \emph{induced} subgraph.
Let $d(v)$ be the degree of node $v$, $\Delta$ be the maximum degree of $G$, and $\Delta_{V^\prime}$ be the maximum degree of the subgraph induced~by~$V^\prime \subseteq V$.
The \emph{graph partitioning} problem~(GP) consists of assigning each node of $G$ to exactly one of $k$ distinct \emph{blocks} respecting a balancing constraint in order to minimize the edge-cut.
More precisely, GPP partitions $V$ into $k$ blocks $V_1$,\ldots,$V_k$ (\ie $V_1\cup\cdots\cup V_k=V$ and $V_i\cap V_j=\emptyset$ for $i\neq j$), which is called a \emph{\mbox{$k$-partition}} of $G$.
The \emph{balancing constraint} demands that the sum of node weights in each block does not exceed a threshold associated with some allowed \emph{imbalance}~$\epsilon$. 
More specifically, $\forall i~\in~\{1,\ldots,k\} \gilt$ $c(V_i)\leq L_{\max}\Is \big\lceil(1+\epsilon) \frac{c(V)}{k} \big\rceil$.

The \emph{edge-cut} of a $k$-partition consists of the total weight of the edges crossing blocks, \ie $\sum_{i<j}\omega(E_{ij})$, where $E_{ij}\Is\setGilt{\set{u,v}\in E}{u\in V_i,v\in V_j}$. 
An abstract view of the partitioned graph is a \emph{quotient graph} $\mathcal{Q}$, in which nodes represent blocks and edges are induced by the connectivity between blocks. 
More precisely, each node $i$  of $\mathcal{Q}$ has weight $c(V_i)$ and there is an edge between $i$ and $j$ if there is at least one edge in the original partitioned graph that runs between the blocks $V_i$ and $V_j$.
We call \emph{neighboring blocks} a pair of blocks that is connected by an edge in the quotient graph.
A vertex $v \in V_i$ that has a neighbor $w \in V_j, i\neq j$, is a boundary vertex.
A successful heuristic for partitioning large graphs is the \emph{multilevel graph partitioning} (MGP) approach depicted in Figure~\ref{fig:mgp}, where the graph is recursively \emph{contracted} to achieve smaller graphs which should reflect the same basic structure as the input graph. 
After applying an \emph{initial partitioning} algorithm to the smallest graph, the contraction is undone and, at each level, a
\emph{local search} method is used to improve the partitioning induced by the coarser level. 
\emph{Contracting} a cluster of nodes $C=\set{u_1, \ldots, u_{\ell}}$ consists of replacing these nodes  by a new node $v$. 
The weight of this node is set to the sum of the weight of the cluster vertices. 
Moreover, the new node is connected to all elements $w \in \bigcup_{i=1}^{\ell} N(u_i)$, $\omega(\set{v,w}) = \sum_{i=1}^{\ell} \omega(\set{u_i,w})$. 
This ensures that a partition from a coarser level that is transfered to a finer level maintains the edge cut and the balance of the partition.
The  \emph{uncontraction} of a node undoes the contraction. Local search moves vertices between the blocks in order to reduce the~objective.


\textbf{Computational Models.}
The focus of this paper is to engineer a graph partitioning algorithm for a streaming input.
In particular, the input is a stream of nodes alongside with their respective adjacency lists.
The classic streaming model is the one-pass model, in which the nodes have to be permanently assigned to a block as soon as they are loaded from the input. 
As soon as assignment decisions of an algorithm for the current node depend on the previous decisions, an algorithm in the model has to store the assignment of the previous loaded nodes and hence needs $\Omega(n)$ space.
We use an extended version of this model, which is called the \emph{buffered streaming} model.
More precisely, a $\delta$-sized \emph{buffer} or \emph{batch} of input nodes with their neighborhood is repeatedly loaded.
Partition/block assignment decisions have to be made after the whole buffer is loaded. 
While we investigate the dependence of our algorithm on this parameter, in practice the parameter  will depend on the amount of available memory on a machine. The parameter can be dynamically chosen such that the buffer is ``full'' if $\Theta(n)$ space has been loaded from the disk.
Hence the buffered streaming model asymptotically does not need more space than a one-pass streaming algorithm if this setting is used. 
This holds true even in the worst case: when a node has degree close to $n$.
In this paper, we use a constant $\delta$ throughout the run of an algorithm.
For a predefined batch size $\delta$, the total amount of $\lceil n/\delta \rceil$ batches are consecutively loaded and assigned to blocks.

\subsection{Related Work}
\label{subsec:related_work}

There has been a large amount of research on graph partitioning. We refer the reader to~\cite{GPOverviewBook,SPPGPOverviewPaper,DBLP:reference/bdt/0003S19} for extensive material and references. Here, we focus on results close to our main contribution.
The most successful general-purpose offline algorithms to solve the graph partitioning problem for huge real-world graphs are based on the multilevel approach. 
The most commonly used formulation of the multilevel scheme for graph partitioning was proposed in~\cite{Hendrickson95}.
However, these algorithms require the graph to fit into main memory  of a single machine or into the memory of a distributed machine if a distributed memory parallel partitioning algorithm is used. 

Stanton and Kliot~\cite{stanton2012streaming} introduced graph partitioning in the streaming model and proposed many natural heuristics to solve it.
These heuristics include one-pass methods such as \AlgName{Hashing}, \AlgName{Chunking}, and \AlgName{linear deterministic greedy} (\AlgName{LDG}), and some buffered methods such as \AlgName{greedy evocut}.
The evocut buffered model is different from our model as it is an extended one-pass model in which a buffer of fixed size is kept and the algorithm can assign any node from the buffer to a block, rather than the one that has been received most recently. However, all methods that use a buffer perform significantly worse than random partitionings -- hence we do not include them in our experiments.
In their experiments, \AlgName{LDG} had the best overall results in terms of total edge-cut.
In this algorithm, node assignments prioritize blocks containing more neighbors while using a penalty multiplier to control imbalance. 
In particular, it assigns a node $v$ to the block $V_i$ that maximizes $|V_i \cap N(v)|*\lambda(i)$ with $\lambda(i)$ being a multiplicative penalty defined as $(1-\frac{|V_i|}{L_\text{max}})$. 
The intuition here is that the penalty avoids to overload blocks that are already~very~heavy.
Tsourakakis~et~al.~\cite{tsourakakis2014fennel} proposed a one-pass partitioning heuristic named \AlgName{Fennel}, which is an adaptation of the widely-known clustering objective \emph{modularity}~\cite{brandes2007modularity}.
Roughly speaking, \AlgName{Fennel} assigns a node $v$ to a block $V_i$, respecting a balancing threshold, in order to maximize an expression of type $|V_i\cap N(v)|-f(|V_i|)$, \ie with an additive penalty.
The assignment decision of \AlgName{Fennel} is based on an interpolation of two properties: attraction to blocks with more neighbors and repulsion from blocks with more non-neighbors.
When $f(|V_i|)$ is a constant, the resulting objective function coincides with the first property.
If $f(|V_i|) = |V_i|$, the objective function coincides with the second property.
More specifically, the authors defined the \AlgName{Fennel} objective function by using $f(|V_i|) = \alpha \times \gamma \times |V_i|^{\gamma-1}$, in which $\gamma$ is a free parameter and $\alpha = m \frac{k^{\gamma-1}}{n^{\gamma}}$.
After a parameter tuning made by the authors, \AlgName{Fennel} uses $\gamma=\frac{3}{2}$, which provides $\alpha=\sqrt{k}\frac{m}{n^{3/2}}$. Note that in the original paper, the authors assume $k$ to be constant and hence derive a complexity of $O(n+m)$. However, since one has to iterate over all blocks $k$ for each node the complexity of the algorithm depends on $k$ and is given by $O(nk+m)$.

A restreaming approach has been introduced by Nishimura~and~Ugander~\cite{nishimura2013restreaming}.
In this model, multiple passes through the entire input are allowed, to allow iterative improvements.
The authors proposed easily implementable restreaming versions of \AlgName{LDG} and \AlgName{Fennel}: \AlgName{ReLDG} and \AlgName{ReFennel}. 
Awadelkarim and Ugander~\cite{awadelkarim2020prioritized} studied the effect of node ordering for streaming graph partitioning.
The authors introduced the notion of \emph{prioritized streaming}, in which (re)streamed nodes are statically or dynamically reordered based on some priority. 
Patwary~\etal~\cite{patwary2019window} proposed WStream, a greedy stream graph partitioning algorithm that keeps a sliding stream window.
This sliding window contains a few hundred nodes such that it gives more information about the next node to be allocated to a block. However, the code is not available and the algorithm has only been evaluated on graphs with a few thousand vertices. 
Jafari~\etal~\cite{jafari2021fast} proposed a shared-memory partitioning algorithm based on a buffered streaming computational model similar to the one we use here.
Their algorithm uses the idea of multilevel algorithm but with a simplified structure in which the \AlgName{LDG} one-pass algorithm constitutes the coarsening step, the initial partitioning, and the local search during uncoarsening.
Our work differs from theirs in the fact that we focus on single-threaded execution,  we construct a sophisticated model instead of processing the nodes of a batch directly, and our algorithm is inspired on the \AlgName{Fennel} one-pass algorithm, which outperformed \AlgName{LDG} in previously published studies.
There are also a wide range of algorithms that focus on (buffered) streaming edge partitioning \cite{petroni2015hdrf, mayer2018adwise, sajjad2016boosting}.
However, as our focus is on edge cut-based algorithms they are  beyond the scope of this work.

\section{Buffered Graph Partitioning}
\label{sec:Streaming Graph Partitioning}

We now present our main contribution, namely \AlgName{HeiStream}: a novel algorithm to solve graph partitioning in the buffered streaming model. 
We start by first outlining the overall structure behind \AlgName{HeiStream} and then we present each of the algorithm components.

\subsection{Overall Structure}
\label{subsec:Overall Structure}
\begin{figure}[t]
	\centering
	\includegraphics[width=0.9\textwidth]{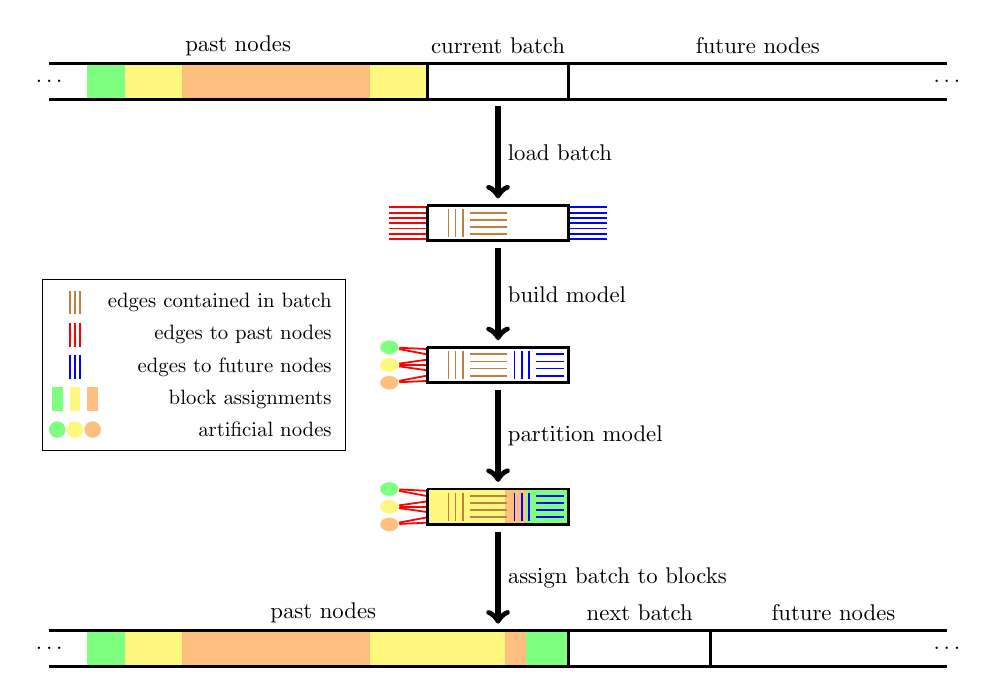}
	\caption{Detailed structure of \AlgName{HeiStream}. The algorithm starts by loading a batch of nodes alongside with their edges. Next, it builds a meaningful model based on the loaded nodes and edges. This model is then partitioned using a multilevel algorithm. Based on the partition of the model, the nodes of the current batch are permanently assigned to blocks. This whole process is repeated for the next batch until the whole graph is partitioned.}
	\label{fig:comprehensive_scheme}
\end{figure}

We now explain the overall structure of \AlgName{HeiStream}.
We slide through the streamed graph $G$ by repeating the following successive operations until all the nodes of $G$ are assigned to blocks.
First, we load a batch containing $\delta$ nodes alongside with their adjacency lists.
Second, we build a model $\mathcal{B}$ to be partitioned. This model represents the already partitioned vertices as well as the nodes of the current batch.
Third, we partition $\mathcal{B}$ with a multilevel partitioning algorithm to optimize for the \AlgName{Fennel} objective function.
And finally, we permanently assign the nodes from the current batch to blocks.
Algorithm~\ref{alg:overall_algorithm} summarizes the general structure of \AlgName{HeiStream} and Figure~\ref{fig:comprehensive_scheme} shows the detailed structure of \AlgName{HeiStream}.

\begin{algorithm}[t] 
	\caption{Structure of \AlgName{HeiStream}} 
	\label{alg:overall_algorithm} 
	\begin{algorithmic} 
		\WHILE{$G$ has not been completely streamed}
		\STATE Load batch of vertices
		\STATE Build model $\mathcal{B}$
		\STATE Run multilevel partitioning on model $\mathcal{B}$
		\STATE Assign nodes of batch to permanent blocks
		\ENDWHILE
	\end{algorithmic}
\end{algorithm}

\subsection{Model Construction}
\label{subsec:Graph Construction}
We build two different models, which yield a running time--quality trade-off. 
We start by describing the \emph{basic model} and then extend this later.
When a batch is loaded, we build the model $\mathcal{B}$ as follows.
We initialize $\mathcal{B}$ as the subgraph of $G$ induced by the nodes of the current batch.
If the current batch is not the first one, we add $k$ artificial nodes to the model.
These represent the $k$ preliminary blocks in their current state, i.e., filled with the nodes from the previous batches, which were already assigned.
The weight of an artificial node $i$ is set to the weight of block $V_i$.
A node of the current batch is connected to an artificial node $i$ if it has a neighbor from the previous batch that has been assigned to block $V_i$.
If this creates parallel edges, we replace them by a single edge with its weight set to the sum of the weight of the parallel edges. 
Note that the basic model does ignore edges towards nodes that will be streamed in future batches, \ie batches that have not been streamed yet.
Our \emph{extended model} incorporates edges towards nodes from future, not yet loaded, batches -- if the stream contains such edges.
We call edges towards nodes of future batches ghost edges and the corresponding endpoint in the future batch ghost node.
Ghost nodes and ghost edges provide partial information about parts of $G$ that have not yet been streamed.
Hence, representing them in the model $\mathcal{B}$ enhances the partitioning process.
Note though that simply inserting all ghost nodes and edges can overload memory in case there is an excessive amount of them.
Thus our approach consists of randomly contracting the ghost nodes with one of their neighboring nodes from the current batch. 
Note that this contraction increments the weight of a node within our model and ensures that if there are more than one node from the current batch connected to the same future node, then there will be edges between those nodes in our model. 
Also note that the contraction ensures that the number of nodes in all models throughout the batched streaming process is constant. 
This prevents memory from being overloaded and makes it unnecessary to reallocate memory for $\mathcal{B}$ between successive batches.
In order to give a lower priority to ghost edges in comparison to the other edges, we divide the weight of each ghost edge by 2 in our~model.
The construction of the extended model is conceptually illustrated in Figure~\ref{fig:comprehensive_scheme}.

\subsection{Multilevel Weighted Fennel}
\label{subsec:Multilevel Fennel}
Our approach to partition the model $\mathcal{B}$ is a multilevel version of the algorithm \AlgName{Fennel}.
Recall that the multilevel scheme consists of three successive phases: coarsening, initial partitioning, and uncoarsening, as depicted in Figure~\ref{fig:mgp}. 
Note, however, that the artificial nodes in our model can become very heavy and are not allowed to be contracted or change their block. As a consequence, these nodes need special handling in our multilevel algorithm.
Moreover, the \AlgName{Fennel} algorithm is designed for unweighted graphs, hence it needs some adaptation to be used in \AlgName{HeiStream}. We introduce a generalization of \AlgName{Fennel} for weighted graphs that can be directly employed in a multilevel algorithm.
In this section, we present this generalized \AlgName{Fennel} objective and explain the details of our multilevel algorithm to partition the model.

\subsubsection{Generalized Fennel.}
As already mentioned, adaptations are necessary to implement \AlgName{Fennel} for weighted graphs, in particular in a multilevel scheme.
First, note that the original formulation of \AlgName{Fennel} only works for unweighted graphs~\cite{tsourakakis2014fennel}.
However, our model $\mathcal{B}$ has nodes and edges that are weighted -- due to connections to the artificial nodes as well as future nodes that may be contracted into the model. Moreover, the multilevel scheme creates a sequence of weighted graphs. 
Note that a generalization of the \AlgName{Fennel} gain function has to ensure that the gain of a node on a coarser level corresponds to the sum of the gains of the nodes that it represents on the finest level. This way the algorithm gets a global view onto the optimization problem on the coarser levels and a very local view on the finer levels. Moreover, it is ensured that on each level of the hierarchy the algorithm works towards optimizing the given objective.

Our generalization of the gain function of \AlgName{Fennel} is as follows. 
Let $u$ be the node that should be assigned to a block to another block. 
Our generalized \AlgName{Fennel} assigns $u$ to a block $i$ that maximizes $\sum_{ v \in V_i \cap N(u)}{ \omega(u,v)} - c(u) f(c(V_i))$, such that $f(c(V_i)) =  \alpha * \gamma * c(V_i)^{\gamma-1}$.
Note that this is a direct generalization of the unweighted case. First, if the graph does not have edge weights, then the first part of the equation becomes $|V_i \cap N(u)|$ which is the first part of the \AlgName{Fennel} objective. Second, if the graph also does not have node weights, then the second part of the equation is the same as the second part of the equation in the \AlgName{Fennel} objective.
Moreover, observe that the penalty term $f(c(V_i))$ in our objective is multiplied by $c(u)$.
This is done to have the property stated above and formalized in Theorem~\ref{theo:generalized_fennel}.
Finally, we keep the original value of the parameters $\alpha$ and $\gamma$ in order to keep consistency. 

\begin{theorem}
	If a set of nodes $S \subseteq V$ is contracted into a node $w$, the generalized \AlgName{Fennel} gain function of $w$ is equal to the sum of the generalized \AlgName{Fennel} gain functions of the nodes in $S$.
	\label{theo:generalized_fennel}
\end{theorem}

\begin{proof}
	On the one hand, the generalized \AlgName{Fennel} gain of assigning a node $w$ to a block~$i$ is $\sum\limits_{ v \in V_i \cap N(w)}{ \omega(w,v)} - c(w) f(c(V_i))$.
	On the other hand, the sum of the generalized \AlgName{Fennel} gains of assigning all nodes in $S$ to a block~$i$ can be expressed as $\sum\limits_{u\in S}\sum\limits_{ v \in V_i \cap N(u)}{ \omega(u,v)} - \sum\limits_{u\in S}c(u) f(c(V_i))$.
	Since the factor $f(c(V_i))$ is identical in both expressions, the two of them are equivalent if the following equalities hold: 
	$\sum\limits_{ v \in V_i \cap N(w)}{ \omega(w,v)} := \sum\limits_{u\in S}\sum\limits_{ v \in V_i \cap N(u)}{ \omega(u,v)}$ and $c(w) := \sum\limits_{u\in S}c(u)$.
	Both of these equalities are trivially true as a property of the contraction process.
\end{proof}

\subsubsection{Multilevel Fennel.}
We now explain our multilevel algorithm to partition the model $\mathcal{B}$.
\label{subsec:Contraction}
Our \emph{coarsening} phase is an adapted version of the size-constraint label propagation approach \cite{pcomplexnetworksviacluster}. To be self-contained, we shortly outline the coarsening approach and then show how to modify it to be able to handle artificial nodes. To compute a graph hierarchy, the algorithm computes a size-constrained clustering on each level and contract that to obtain the next level.
The clustering is contracted by replacing each cluster by a single node, and the process is repeated recursively until the graph becomes small enough.
This hierarchy is then used by the algorithm.
Due to the way we define contraction, it ensures that a partition of a coarse graph corresponds to a partition of all the finer graphs in the hierarchy with the same edge-cut and balance. 
Note that cluster contraction is an aggressive coarsening strategy. 
In contrast to matching-based approaches, it enables us to drastically shrink the size of irregular networks.
The intuition behind this technique is that a clustering of the graph (one hopes) contains many edges running inside the clusters and only a few edges running between clusters, which is favorable for the~edge~cut~objective.

The algorithm to compute clusters is based on \emph{label propagation}~\cite{labelpropagationclustering} and avoids large clusters by using a \emph{size constraint}, as described in~\cite{pcomplexnetworksviacluster}. 
For a graph with $n$ nodes and $m$ edges, one round of size-constrained label propagation can be implemented to run in $O(n+m)$ time.
Initially, each node is in its own cluster/block, \ie the initial block ID of a node is set to its node ID.
The algorithm then works in rounds. 
In each round, all the nodes of the graph are traversed. 
When a node $v$ is visited, it is \emph{moved} to the block that has the strongest connection to $v$, \ie it is moved to the cluster $V_i$ that maximizes $\omega(\{(v, u) \mid u \in N(v) \cap V_i \})$. 
Ties are broken randomly. 
We perform at most $L$ rounds, where $L$ is a tuning~parameter.

In \AlgName{HeiStream}, we have to ensure that two artificial nodes are not contracted together since each of them should remain in its previously assigned block. 
We achieve this by ignoring artificial nodes and artificial edges during the label propagation, \ie artificial nodes cannot not change their label and nodes from the batch can not change their label to become a label of an artificial node.  As a consequence, artificial nodes are not contracted during coarsening.
Overall, we repeat the process of computing a size-constrained clustering and contracting it, recursively. 
As soon as the graph is small enough, \ie it has fewer nodes than an $O(\max({|\mathcal{B}|/k,k}))$ threshold, it is initially partitioned by an initial partitioning~algorithm. 
More precisely, we use the threshold $\max({\frac{|\mathcal{B}|}{2xk},xk})$, in which $x$ is a tuning parameter.
Note that, for large enough buffer sizes, this threshold will~be~$O(|\mathcal{B}|/k)$.

\label{subsec:Initial Partition}

\label{subsec:Uncoarsening}

When the coarsening phase ends, we run an \emph{initial partitioning algorithm} to compute an initial $k$-partition for the coarsest version of $\mathcal{B}$.
That means that all nodes other than the artificial nodes, which are already assigned, will be assigned to blocks. 
To assign the nodes, we run our generalized \AlgName{Fennel} algorithm with explicit balancing constraint $L_{\max}$, \ie the weight of no block will exceed $L_{\max}$.
To be precise, a node $u$ will be assigned to a block $i$ that maximizes $\sum_{ v \in V_i \cap N(u)}{ \omega(u,v)} - c(u) f(c(V_i))$, such that $f(c(V_i)) =  \alpha * \gamma * c(V_i)^{\gamma-1}$ \emph{and} $c(V_i \cup u) \leq L_{\max}$. 
Note that the algorithm at this point considers all possible blocks $i \in \{1, \ldots, k\}$ and hence has complexity proportional to~$k$. 
However, as the coarsest graph has $O(|\mathcal{B}|/k)$ nodes, overall the initial partitioning needs time which is linear in the size of the~input~model.
When initial partitioning is done, we transfer the current solution to the next finer level by assigning each node of the finer level to the block of its coarse representative. 
At each level of the hierarchy, we apply a \emph{local search} algorithm.
Our local search algorithm is the same size-constraint label propagation algorithm we used in the contraction phase but with a different objective function.
Namely, when visiting a node node, we remove it from its current block and then we assign it to the neighboring block which maximizes the generalized \AlgName{Fennel} gain function defined above.
Note that, in contrast to the initial partitioning, only blocks of adjacent nodes are considered here. 
Hence, one round of the algorithm can still be implemented to run in linear  time in the size the current level.
As in the coarsening phase, artificial nodes cannot be moved between blocks.
Differently though, we do not exclude the artificial nodes from the label propagation here.
This is the case because the artificial nodes and their edges are used to compute the generalized \AlgName{Fennel} gain function of the other nodes. 
As in the initial partitioning, we use the explicit size constraint $L_{\max}$~of~$G$. As a side note, we also tried to use high-quality offline algorithms as initial partitioning algorithms, however, in preliminary experiments this results in very unbalanced blocks (even with adaptively configured balance constraints) and overall in reduced quality throughout the process. Hence, we did not consider this~further.

Assuming geometrically shrinking graphs thoughout the hierarchy and assuming that the buffer size $\delta$ is larger than the number of blocks $k$, then the overall running time to partition a batch is linear in the size of the batch.
This is due to the fact that the overall running time of coarsening and local search sums up to be linear in the size of the batch, while the overall running time of the initial partitioning depends linearly on the size of the input model. 
Summing this up over all batches yields overall linear running time $O(n+m)$.

\subsection{Restreaming}
\label{subsec:Restreaming}
We now extend \AlgName{HeiStream} to operate in a restreaming setting.
During restreaming, the overall structure of the algorithm is roughly the same.
Nevertheless we need to implement some adaptations which we explain in this section.
The first adaptations concern model construction.
Recall that the nodes from the current batch are already assigned to blocks during the previous pass of the input.
We explicitly assign these nodes to their respective blocks in $\mathcal{B}$.
Furthermore, ghost nodes and edges are not needed to construct $\mathcal{B}$.
This is the case since all nodes from future batches are already known and assigned to blocks, \ie these nodes will be represented in the artificial nodes.
More precisely, we adapt the artificial nodes to represent the nodes from all batches except the current one.
Since a partition of the graph is already given,  we do not allow the contraction of cut edges during restreaming in the coarsening phase of our multilevel scheme.
That means that clusters are only allowed to grow inside blocks.
As a consequence, we can directly use the partition computed in the previous pass as initial partitioning for $\mathcal{B}$ so we do not need to run an initial~partitioning~algorithm.

\subsection{Implementation Details}
\label{subsec:Implementation Details}
Our implementation of $\mathcal{B}$ is based on an adjacency array and consecutive node IDs.
We reserve the first $\delta$ IDs for the nodes from the current batch, which keep their global order.
This means that, when we process the $i^{th}$ batch, nodes IDs can be easily converted from our model $\mathcal{B}$ to $G$ and the other way around by respectively summing or subtracting $(i-1)*\delta$ on their ID.
Similarly, we reserve the last $k$ IDs of $\mathcal{B}$ for the artificial nodes and keep their relative order for all batches.
Note that this configuration separates \emph{mutable} nodes (nodes from current batch) and \emph{immutable} nodes (artificial nodes).
This allows us to efficiently control which nodes are allowed to move during coarsening, initial partitioning, and local search.
We keep an array of size $n$ store the permanent block assignment of the nodes of $G$.
To improve running time, we use approximate computation of powering in our \AlgName{Fennel} function.

\section{Experimental Evaluation}
\label{sec:Experimental Evaluation}

\begin{table}[t]
	\footnotesize
	\centering
	\setlength{\tabcolsep}{3pt}
	\caption{Graphs for experiments.}
	\begin{tabular}[t]{| l  r  r  r | }
		\hline
		Graph & $n$& $m$ & Type\\
		\hline  \hline

		\multicolumn{4}{|c|}{Tuning Set} \\
		\hline

		coAuthorsCiteseer & \numprint{227320}    & \numprint{814134}  & Citations \\
		citationCiteseer & \numprint{268495}    & \numprint{1156647}  & Citations \\
		
		amazon0312 & \numprint{400727}  & \numprint{2349869} & Co-Purch. \\
		amazon0601 & \numprint{403364} & \numprint{2443311} & Co-Purch. \\
		amazon0505 & \numprint{410236} & \numprint{2439437} & Co-Purch. \\

		roadNet-PA & \numprint{1087562} & \numprint{1541514} & Roads \\

		com-Youtube & \numprint{1134890}  & \numprint{2987624} & Social \\
		soc-lastfm & \numprint{1191805} & \numprint{4519330}   & Social \\
		
		roadNet-TX & \numprint{1351137} & \numprint{1879201} & Roads \\

		in-2004 & \numprint{1382908}    & \numprint{13591473}  & Web \\
		
		G3\_circuit & \numprint{1585478}    & \numprint{3037674}  & Circuit \\
		
		soc-pokec & \numprint{1632803} & \numprint{22301964}   & Social \\
		
		as-Skitter & \numprint{1694616}    & \numprint{11094209}  & Aut.Syst. \\
		
		wiki-topcats & \numprint{1791489} & \numprint{28511807} & Social \\
		
		roadNet-CA & \numprint{1957027}  & \numprint{2760388} & Roads \\

		wiki-Talk & \numprint{2388953}     & \numprint{4656682}  & Web \\
		
		soc-flixster & \numprint{2523386} & \numprint{7918801}   & Social \\

		del22 & \numprint{4194304}  & \numprint{12582869} & Artificial \\
		rgg22 & \numprint{4194304} & \numprint{30359198} & Artificial \\
		
		del23 & \numprint{8388608}  & \numprint{25165784} & Artificial \\
		rgg23 & \numprint{8388608} & \numprint{63501393} & Artificial \\
		\hline

		\multicolumn{4}{|c|}{Huge Graphs} \\
		\hline

		uk-2005 &   \numprint{39459923} & \numprint{783027125} & Web \\
		twitter7 & \numprint{41652230} & \numprint{1202513046} & Social \\
		sk-2005 & \numprint{50636154} & \numprint{1810063330} & Web \\
		soc-friendster & \numprint{65608366}  & \numprint{1806067135} & Social \\
		er-fact1.5s26 & \numprint{67108864}  & \numprint{907090182} & Artificial \\
		RHG1 & \numprint{100000000}  & \numprint{1000913106} & Artificial \\
		RHG2 & \numprint{100000000}  & \numprint{1999544833} & Artificial \\
		uk-2007-05 & \numprint{105896555} & \numprint{3301876564}   & Web \\
		\hline
		\end{tabular}
		\begin{tabular}[t]{| l  r  r  r | }
		\hline
		Graph & $n$& $m$ & Type\\
		\hline  \hline	
		
		\multicolumn{4}{|c|}{Test Set} \\
		\hline

		Dubcova1 & \numprint{16129} & \numprint{118440} & Meshes \\
		hcircuit & \numprint{105676}  & \numprint{203734} & Circuit \\

		coAuthorsDBLP & \numprint{299067}     & \numprint{977676}  & Citations \\
		Web-NotreDame & \numprint{325729}     & \numprint{1090108}  & Web \\
		Dblp-2010 & \numprint{326186}     & \numprint{807700}  & Citations \\
		ML\_Laplace & \numprint{377002} & \numprint{13656485} & Meshes \\
		coPapersCiteseer & \numprint{434102}     & \numprint{16036720}  & Citations \\
		coPapersDBLP & \numprint{540486}     & \numprint{15245729}  & Citations \\
		
		Amazon-2008 & \numprint{735323}  & \numprint{3523472} & Similarity \\
		eu-2005 & \numprint{862664}    & \numprint{16138468}  & Web \\
		web-Google & \numprint{916428}    & \numprint{4322051}  & Web \\

		ca-hollywood-2009 & \numprint{1087562} & \numprint{1541514} & Roads \\
		
		Flan\_1565 & \numprint{1564794} & \numprint{57920625} & Meshes \\
		
		Ljournal-2008 & \numprint{1957027}  & \numprint{2760388} & Social \\
		
		HV15R & \numprint{2017169}  & \numprint{162357569} & Meshes \\
		Bump\_2911 & \numprint{2911419}  & \numprint{62409240} & Meshes \\
		del21 & \numprint{2097152}  & \numprint{6291408} & Artificial \\	
		rgg21 & \numprint{2097152} & \numprint{14487995} & Artificial \\
		
		FullChip & \numprint{2987012} & \numprint{11817567} & Circuit \\
		soc-orkut-dir & \numprint{3072441} & \numprint{117185083} & Social \\
		patents & \numprint{3750822}     & \numprint{14970766}  & Citations \\
		cit-Patents & \numprint{3774768}     & \numprint{16518947}  & Citations \\
		soc-LiveJournal1 & \numprint{4847571} & \numprint{42851237}   & Social \\
		circuit5M & \numprint{5558326} & \numprint{26983926} & Circuit \\
		italy-osm & \numprint{6686493}  & \numprint{7013978} & Roads \\
		great-britain-osm & \numprint{7733822} & \numprint{8156517} & Roads \\

		\hline
		
		\end{tabular}

	\label{tab:graphs}
	\vspace*{-0.25cm}
\end{table}

\textbf{Methodology.}  We performed the implementation of \AlgName{HeiStream} and competing algorithms inside the \AlgName{KaHIP} framework (using C++) and compiled them using gcc 9.3 with full optimization turned on (-O3 flag). 
Since no official versions of the one-pass streaming and restreaming algorithms are available in public repositories, we implemented them in our framework.
Our implementations of these algorithms reproduce the results presented in the respective papers and are optimized for running time as much as possible. 
To this end, we implemented \AlgName{Hashing}, \AlgName{LDG}, \AlgName{Fennel}, and \AlgName{ReFennel}.
\AlgName{Multilevel~LDG}~\cite{jafari2021fast} is also not publicly available. 
We sent a message to the authors requesting an executable version of their algorithm for our tests but we have not receive any response up to the moment this work was submitted.
Hence, we compare \AlgName{HeiStream} against \AlgName{Multilevel~LDG} based on the results explicitly reported in~\cite{jafari2021fast}.
We have used two machines.
Machine A has a two six-core Intel Xeon  E5-2630 processor running at $2.8$ GHz, $64$ GB of main memory, and $3$ MB of L2-Cache. 
It runs Ubuntu GNU/Linux 20.04.1 and Linux kernel version 5.4.0-48. 
Machine B has a four-core Intel Xeon E5420 processor running at $2.5$ GHz, $16$ GB of main memory, and $24$ MB of L2-Cache. 
The machine runs Ubuntu GNU/Linux 20.04.1 and Linux kernel version 5.4.0-65. 
Most of our experiments were run on a single core of Machine A.
The only exceptions are the experiments with huge graphs, which were run on a single core of Machine B. When using machine A, we stream the input directly from the internal memory, and when using machine $B$, that only has 16GB of main memory, we stream the input from the~hard~disk.

We use $k \in \{2,3,\ldots,128\}$ for most experiments. 
We allow an imbalance of $3\%$ for all experiments (and all algorithms). All partitions computed by all algorithms have been balanced.
Depending on the focus of the experiment, we measure running time and/or edge-cut.
In general, we perform ten repetitions per algorithm and instance using different random seeds for initialization, and we compute the arithmetic average of the computed objective functions and running time per instance.
When further averaging over multiple instances, we use the geometric mean in order to give every instance the same influence on the \textit{final score}. 
Unless explicitly mentioned otherwise, we average all results of each algorithm grouped by $k$.
For a $k_o$-partition generated by an algorithm $A$, we express its score $\sigma_A$ (which can be edge-cut or running time) using one or more of the following tools:
\emph{improvement} over an algorithm~$B$, computed as $\big(\frac{\sigma_B}{\sigma_A}-1\big)*100\%$;
\emph{ratio}, computed as $\big(\frac{\sigma_A}{\sigma_{max}}\big)$ with $\sigma_{max}$ being the maximum score for $k_o$ among all competitors including $A$;
\emph{relative} value over an algorithm~$B$, computed as $\big(\frac{\sigma_A}{\sigma_{B}}\big)$.
We also present \emph{performance profiles} which  
relate the running time (resp. solution quality) of a group of algorithms to the fastest (resp. best) one on a per-instance basis (rather than grouped by $k$).
Their x-axis shows a factor $\tau$ while their y-axis shows the percentage of instances for which A has up to $\tau$ times the running time (resp. solution quality) of the fastest (resp. best)~algorithm.


\textbf{Instances.}
We get graphs from various sources to test our algorithm \cite{snapnets,nr-aaai15,benchmarksfornetworksanalysis,kappa,funke2019communication}.
Most of the considered graphs were used for benchmark in previous works on graph partitioning.
The graphs wiki-Talk and web-Google, as well as most networks of co-purchasing, roads, social, web, autonomous systems, citations, circuits, similarity, meshes, and miscellaneous are publicly available either in~\cite{snapnets} or in \cite{nr-aaai15}.
Prior to our experiments, we converted these graphs to a vertex-stream format while removing parallel edges, self loops, and directions, and assigning unitary weight to all nodes and edges.
We also use graphs such as eu-2005, in-2004, uk-2002, and uk-2007-05, which are available at the 10$^{th}$ DIMACS Implementation Challenge website~\cite{benchmarksfornetworksanalysis}. 
Finally, we include some artificial random graphs.
We use the name \Id{rggX} for \emph{random geometric graph} with
$2^{X}$ nodes where nodes represent random points in the unit square and edges connect nodes whose Euclidean distance is below $0.55 \sqrt{ \ln n / n }$. 
We use the name \Id{delX} for a graph based on a Delaunay triangulation of $2^{X}$ random points in the unit square~\cite{kappa}.
We use the name \Id{RHGX} for random hyperbolic graphs~\cite{funke2019communication,DBLP:journals/corr/abs-2003-00736} with $10^8$~nodes and $X\times 10^9$~edges.
Basic properties of the graphs under consideration can be found in Table~\ref{tab:graphs}.
For our experiments, we split the graphs in three disjoint sets.
A \emph{tuning} set for the parameter study experiments, a \emph{test} set for the comparisons against the state-of-the-art, and a set of \emph{huge graphs} for special larger~scale~tests. 
In any case, when streaming the graphs we use the natural given order of the nodes.


\subsection{Parameter Study}
\label{subsec:Algorithm Configuration}

\newcommand{ \scaleFactor} {0.62}
\newcommand{ \imgScaleFactor} {0.9}
\newcommand{ \capPosition} {-.45cm}
\newcommand{ \afterCap} {-.45cm}

\begin{figure*}[p]
	\captionsetup[subfigure]{justification=centering}
	\centering
		\vspace*{-.5cm}
	\begin{subfigure}[t]{\scaleFactor\textwidth}
		\centering
		\includegraphics[width=\imgScaleFactor\textwidth]{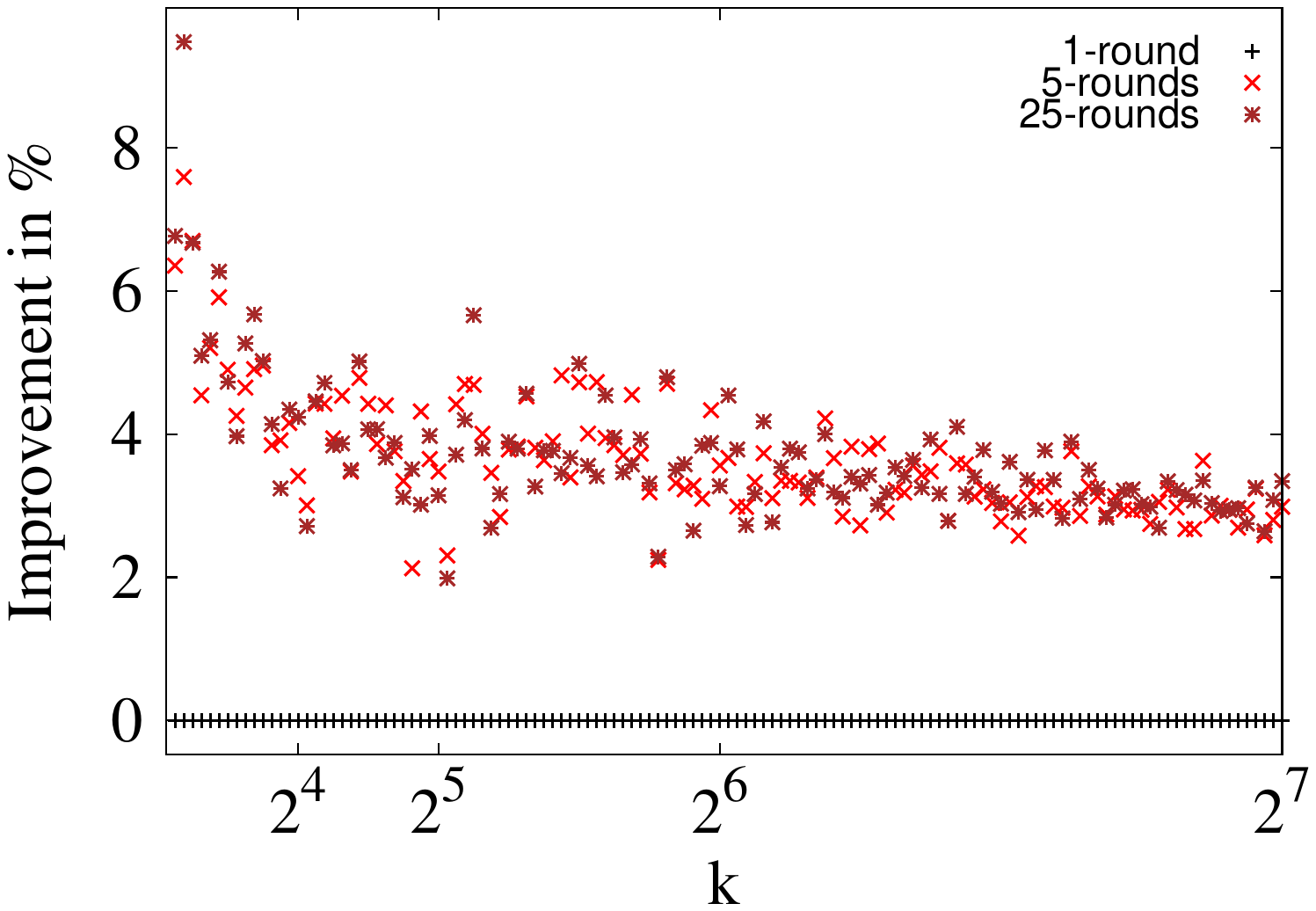}
		\vspace*{\capPosition}
		\caption{Quality improvement plot for label propagation rounds during uncoarsening.}
		\label{fig:res_rep_label}
	\end{subfigure}

	\vspace*{\afterCap}
	\begin{subfigure}[t]{\scaleFactor\textwidth}
		\centering
		\includegraphics[width=\imgScaleFactor\textwidth]{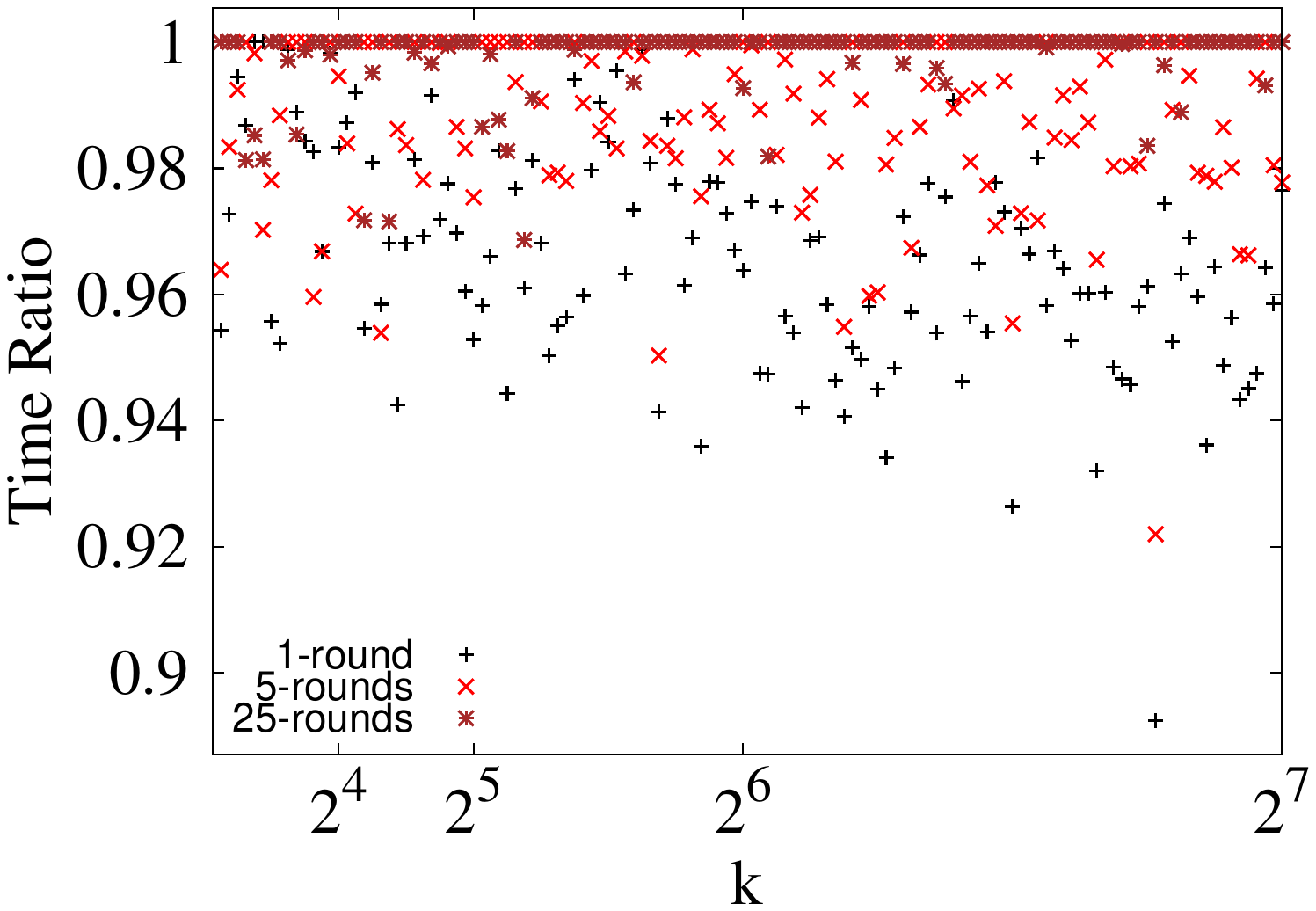}
		\vspace*{\capPosition}
		\caption{Running time ratio plot for label propagation rounds during uncoarsening.}
		\label{fig:tim_rep_label}
	\end{subfigure}
	
	\vspace*{\afterCap}
	\begin{subfigure}[t]{\scaleFactor\textwidth}
		\centering
		\includegraphics[width=\imgScaleFactor\textwidth]{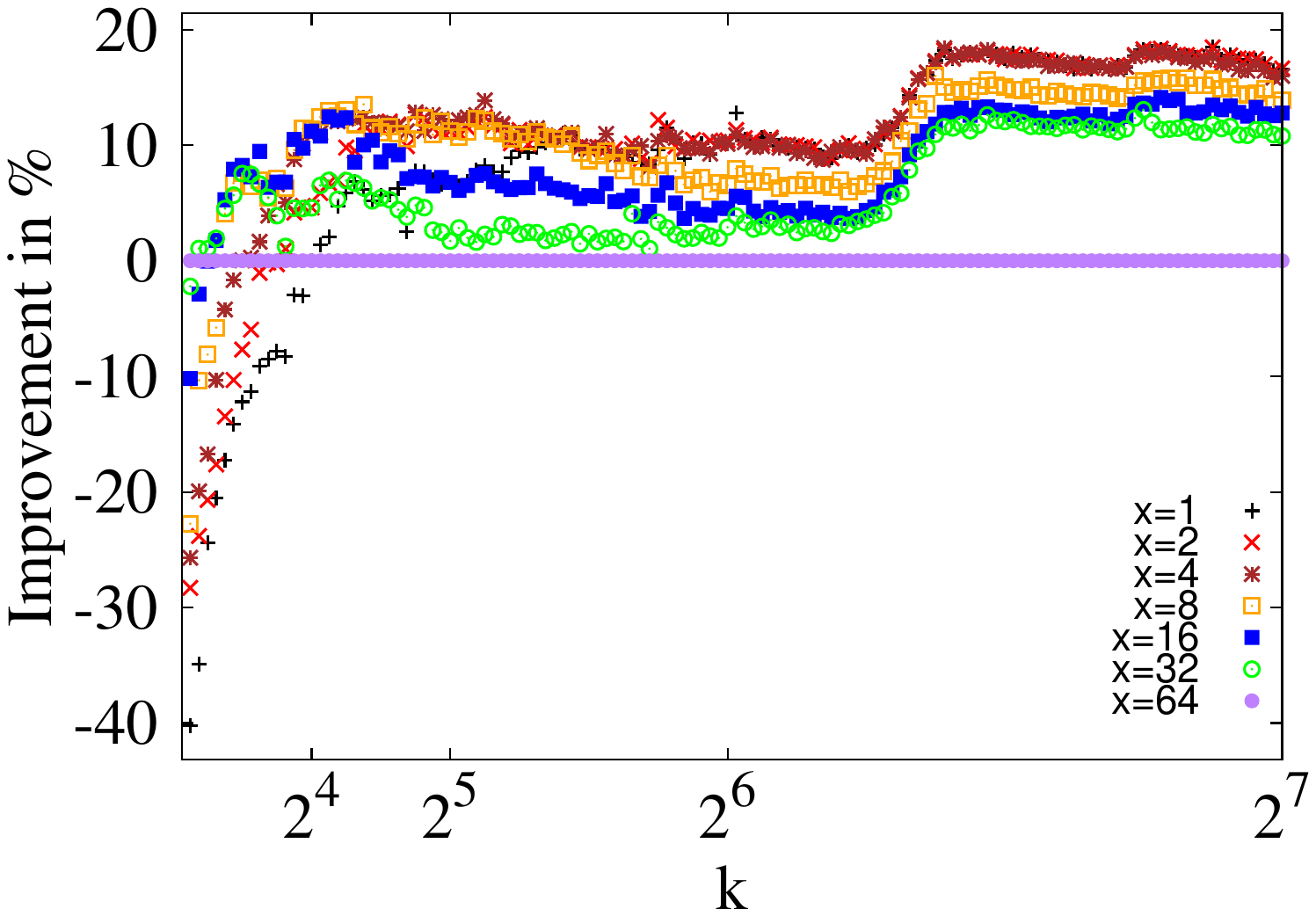}
		\vspace*{\capPosition}
		\caption{Quality improvement plot for parameter $x$ from expression $\max({|\mathcal{B}|/2xk,xk})$.}
		\label{fig:StopRule}
	\end{subfigure}

	\vspace*{.35cm}
	\vspace*{\capPosition}
	\caption{Results for tuning and exploration experiments. Higher is better for quality improvement plots. Lower is better for running time ratio plots.}
	\label{fig:tuning_plots1}
\end{figure*} 

\begin{figure*}[p]
	\captionsetup[subfigure]{justification=centering}
	\centering

	\begin{subfigure}[t]{\scaleFactor\textwidth}
		\centering
		\includegraphics[width=\imgScaleFactor\textwidth]{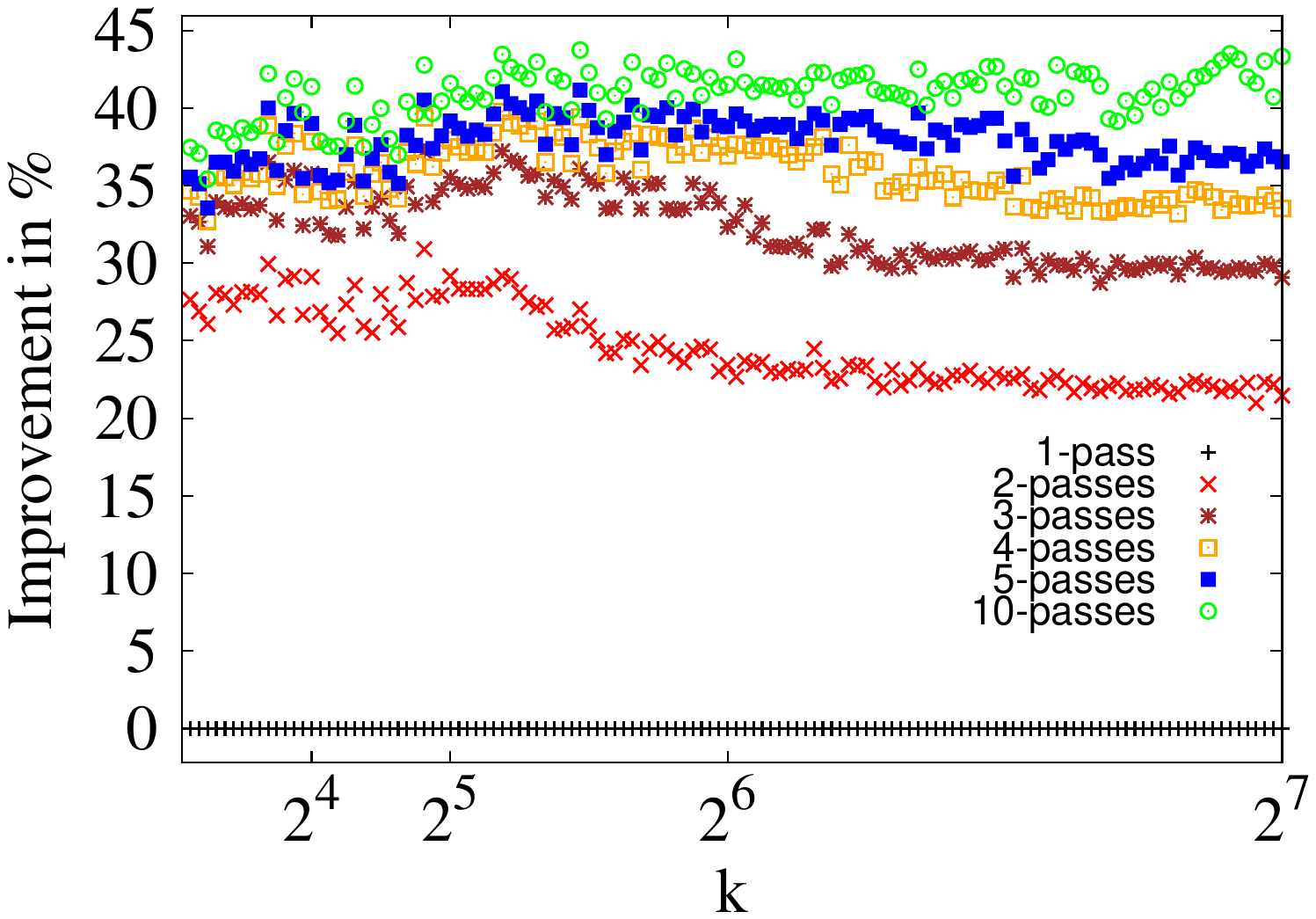}
		\vspace*{\capPosition}
		\caption{Quality improvement plot for restreaming.}
		\label{fig:res_restream}
	\end{subfigure}
	\vspace*{\afterCap}
	\begin{subfigure}[t]{\scaleFactor\textwidth}
		\centering
		\includegraphics[width=\imgScaleFactor\textwidth]{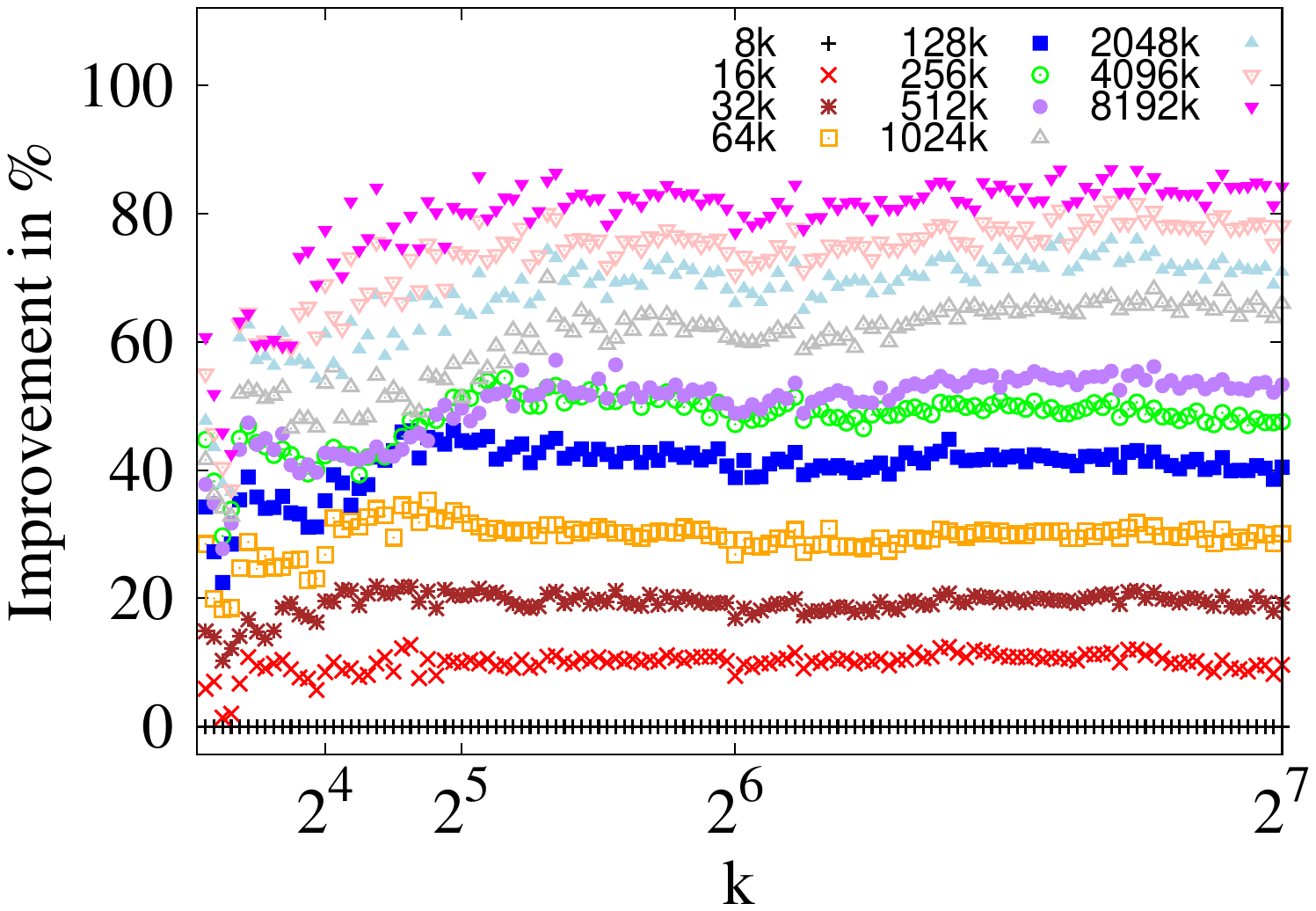}
		\vspace*{\capPosition}
		\caption{Quality improvement plot for buffer size.}
		\label{fig:res_batch_noFennel}
	\end{subfigure}
	\begin{subfigure}[t]{\scaleFactor\textwidth}
		\centering
		\includegraphics[width=\imgScaleFactor\textwidth]{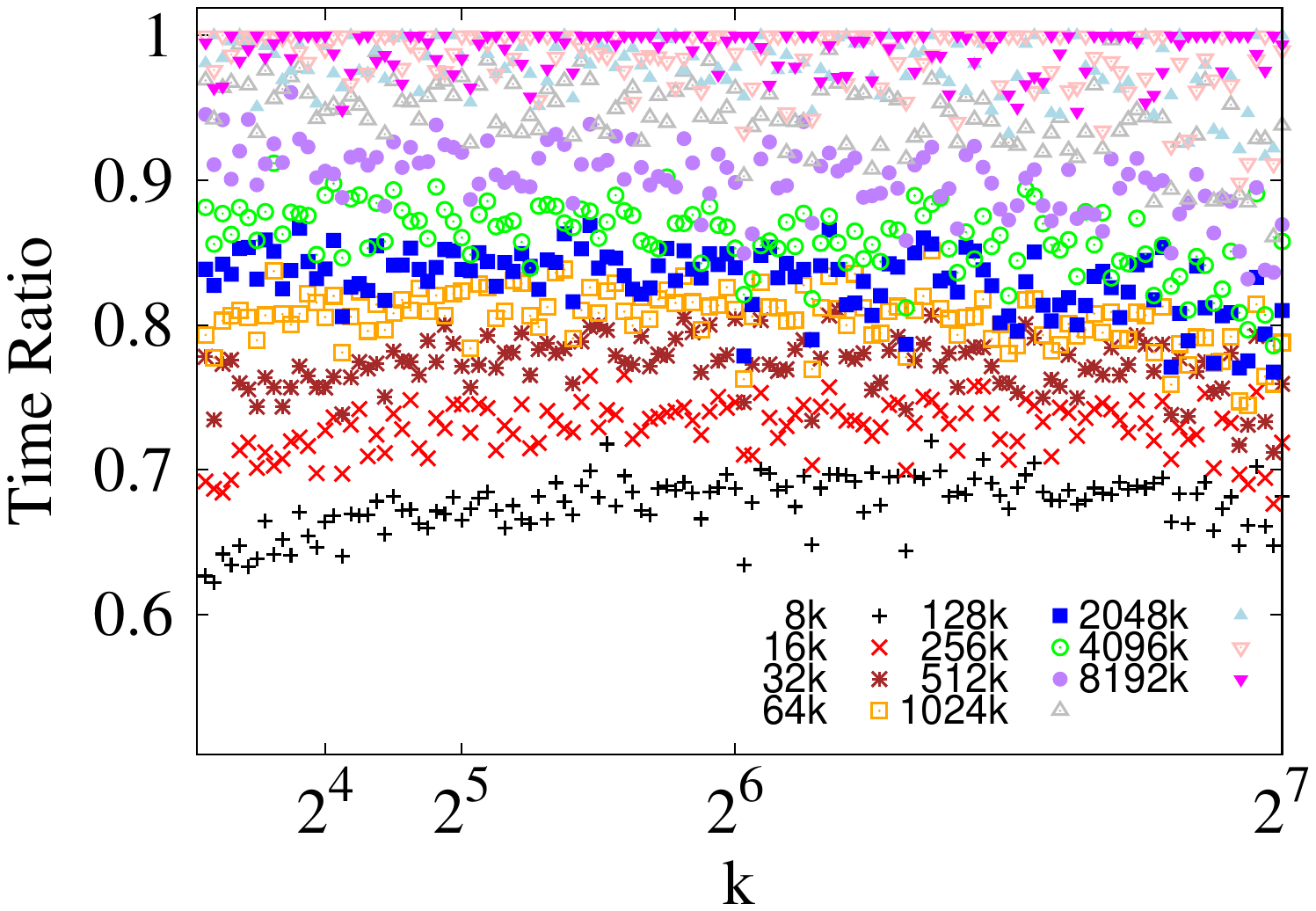}
		\vspace*{\capPosition}
		\caption{Running time ratio plot for buffer size.}
		\label{fig:tim_batch_noFennel}
	\end{subfigure}

	\vspace*{.35cm}
	\vspace*{\capPosition}
	\caption{Results for tuning and exploration experiments. Higher is better for quality improvement plots. Lower is better for running time ratio plots.}
	\label{fig:tuning_plots2}
\end{figure*}

\begin{figure*}[t]
	\captionsetup[subfigure]{justification=centering}
	\centering
	
	\begin{subfigure}[t]{\scaleFactor\textwidth}
		\centering
		\includegraphics[width=\imgScaleFactor\textwidth]{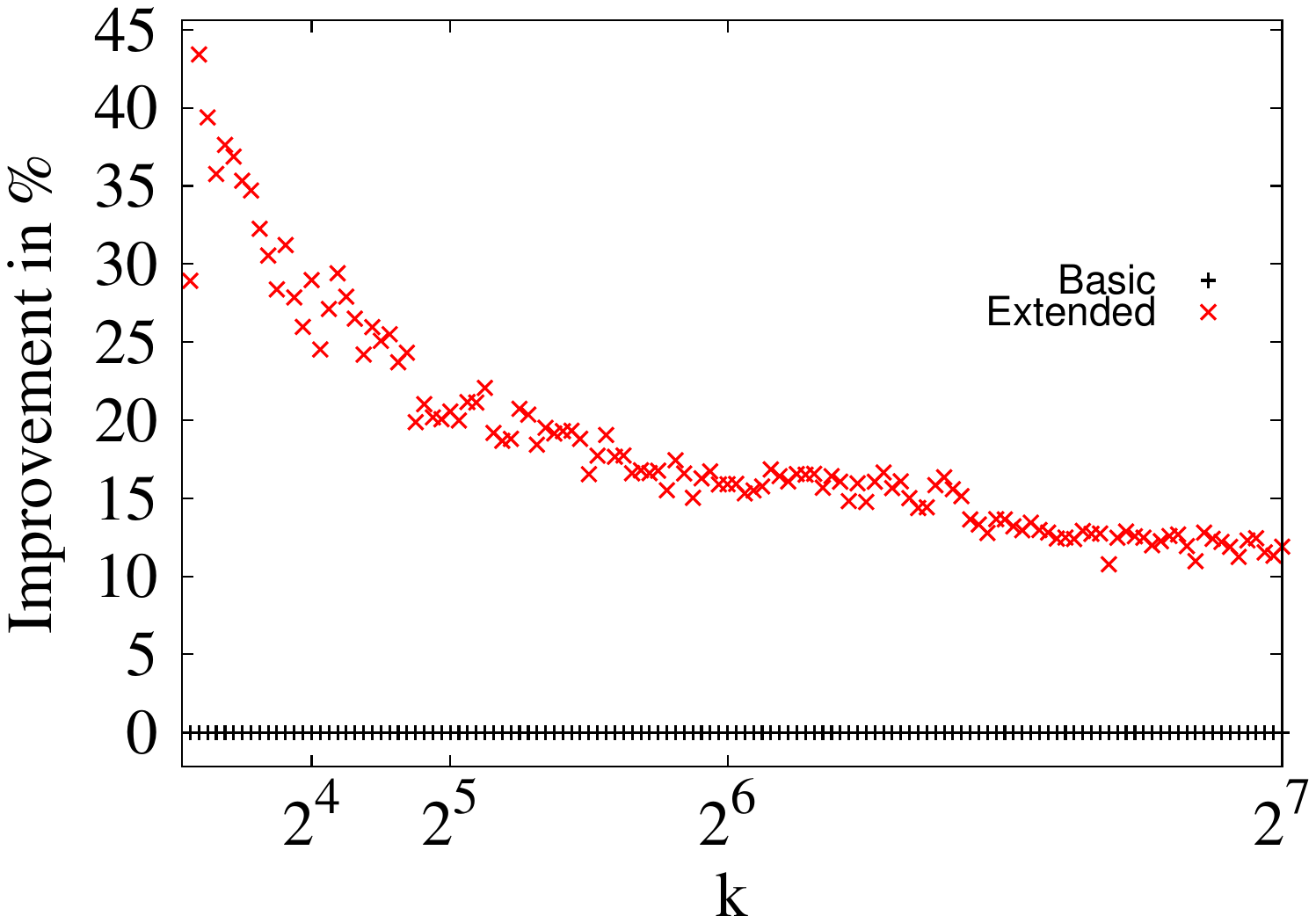}
		\vspace*{\capPosition}
		\caption{Quality improvement plot for model construction.}
		\label{fig:res_ghost}
	\end{subfigure}\hspace{5mm}
	\begin{subfigure}[t]{\scaleFactor\textwidth}
		\centering
		\includegraphics[width=\imgScaleFactor\textwidth]{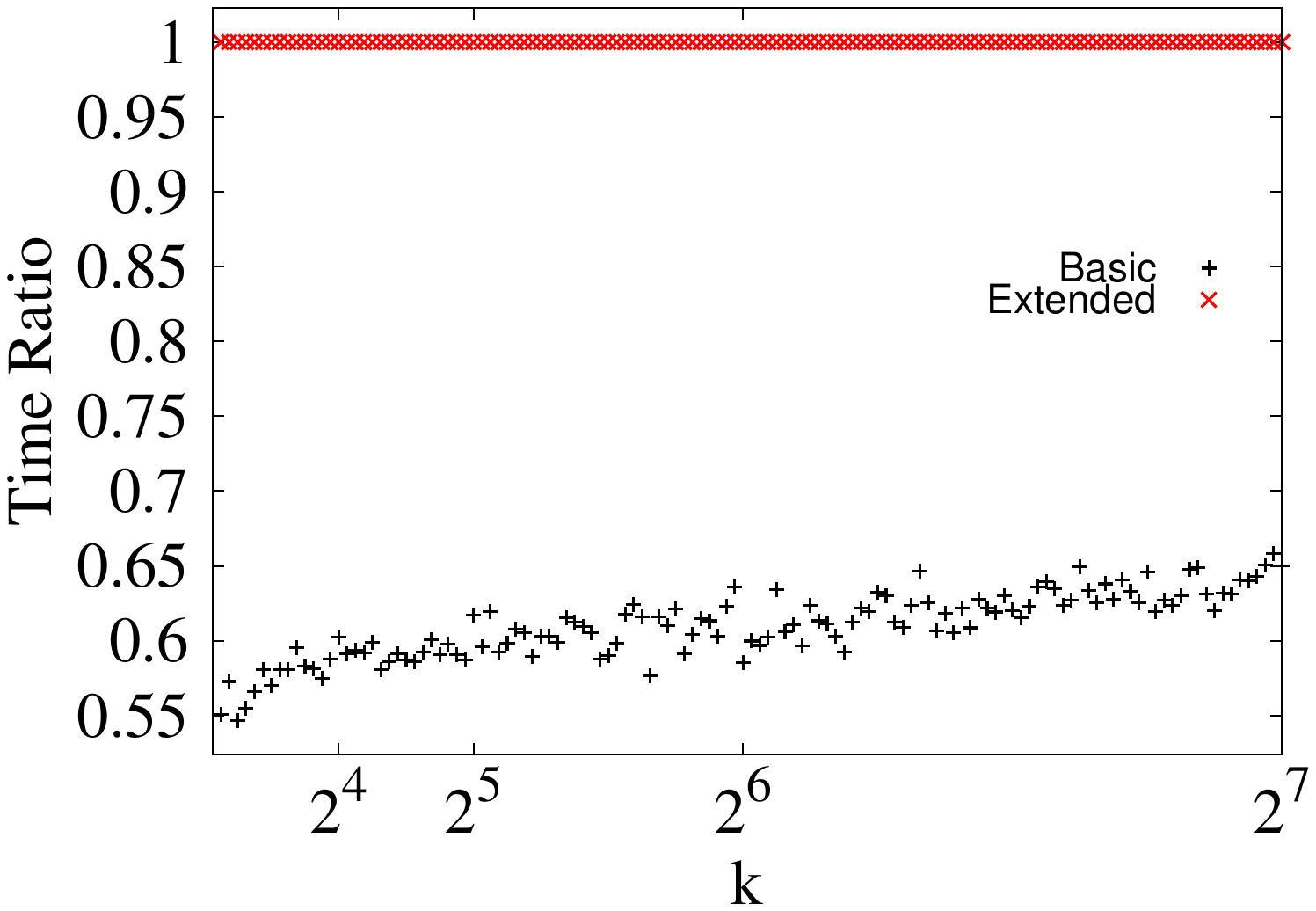}
		\vspace*{\capPosition}
		\caption{Running time ratio plot for model construction.}
		\label{fig:tim_ghost}
	\end{subfigure}

	\vspace*{.35cm}
	\caption{Results for tuning and exploration experiments. Higher is better for quality improvement plots. Lower is better for running time ratio plots.}
	\label{fig:tuning_plots3}

	\vspace*{-1cm}
\end{figure*}

We now present experiments to tune \AlgName{HeiStream} and explore its behavior.
We do this on the tuning set of our instance set.
In our strategy, each experiment focuses on a single parameter of the algorithm while all the other ones are kept invariable.
We start with a baseline configuration consisting of the following parameters:  $5$ rounds in the coarsening label propagation, $1$ round in the local search label propagation, and $x=64$ in the expression of the coarsest model size.
After each tuning experiment, we update the baseline to integrate the best found parameter.
Unless explicitly mentioned otherwise, we run the experiments of this section over all tuning graphs from Table~\ref{tab:graphs} based on the \emph{extended} model construction, \ie including ghost nodes and edges, for a buffer size of $\numprint{32768}$.

\textbf{Tuning.}
We begin by evaluating how the number of label propagation rounds during local search affects running time and solution quality.
In particular, we run configurations of \AlgName{HeiStream} with $1$, $5$, and $25$ rounds and report results in Figures~\ref{fig:res_rep_label}~and~\ref{fig:tim_rep_label}.
Observe that the results of the baseline have considerably lower solution quality than the other ones overall.
On the other hand, the results of the configurations with $5$ and $25$ rounds differ slightly to each other.
On average, they respectively improve solution quality $3.6\%$ and $3.7\%$ over the baseline.
Regarding running time, they respectively increase $2.2\%$ and $3.4\%$ on average over the baseline.
Since the variation of quality for these two configurations is not significant, we decided to integrate the fastest one among them in the algorithm, namely the $5$-round configuration.

Next we look at the parameter $x$ associated with the expression $\max\big({|\mathcal{B}|/2xk,xk}\big)$, which determines the size of the coarsest model.
We run experiments for $x=2^i$, with $i \in \{0,\ldots,6\}$, and report results in Figure~\ref{fig:StopRule}.
We omit running time charts for this experiment since the tested configurations present comparable behavior in this regard.
Figure~\ref{fig:StopRule} shows that the baseline presents the overall worst solution quality while $x=2$ and $x=4$ present the overall best solution quality.
Averaging over all instances, these two latter configurations produce results respectively $10.3\%$ and $11.2\%$ better than the baseline.
In light of that, we decide in favor of $x=4$ to compose \AlgName{HeiStream}.

\textbf{Exploration.}
We start the exploration of open parameters by investigating how the buffer size affects solution quality and running time.
We use as baseline a buffer of $\numprint{8192}$ nodes and successively double its capacity until any graph from the tuning set in Table~\ref{tab:graphs} fits in a single buffer.
We plot our results in Figures~\ref{fig:res_batch_noFennel}~and~\ref{fig:tim_batch_noFennel}.
Note that solution quality and running time increase regularly as the buffer size becomes larger.
This behavior occurs because larger buffers enable more comprehensive and complex graph structures to be exploited by our multilevel algorithm.
As a consequence, there is a trade-off between solution quality and resource consumption.
In other words, we can improve partitioning quality at the cost of considerable extra memory and slightly more running time.
Otherwise, we can save memory as much as possible and get a faster partitioning process at the cost of lowering solution quality.
In practice, it means that \AlgName{HeiStream} can be adjusted to produce partitions as refined as possible with the resources available in a specific system. 
For the extreme case of a single-node buffer, \AlgName{HeiStream} behaves exactly as \AlgName{Fennel}, while it behaves as an internal memory partitioning algorithm for the opposite extreme~case.

Next, we compare the effect of using the \emph{extended} model, which incorporates ghost nodes, over the \emph{basic} model, which ignores ghost nodes.
Figures~\ref{fig:res_ghost}~and~\ref{fig:tim_ghost} displays the results.
The results show that the extended model provides improved quality over the basic model, with an $18.3\%$ improvement on average.
This happens because the presence of ghost nodes and edges expands the perspective of the partitioning algorithm to future batches.
This has a similar effect to increasing the size of the buffer, but at no considerable extra memory cost.
Regarding running time, the results show that the extended model is consistently slower than the basic model for all values of $k$.
Averaging over all instances, the extended model costs $63.9\%$ more running time than the basic model.
This increase in running time is explained by the higher number of edges to be processed when ghost nodes are incorporated in the model.
As a practical conclusion from the experiment, the extended model can be used for better overall partitions with no significant extra memory but at the cost of extra running time.
Otherwise, the basic model can be used for a consistently faster execution at the cost of a lower solution quality.

Finally, we test to what extent solution quality can be improved by restreaming \AlgName{HeiStream} multiple times.
We investigate this by restreaming each input graph $10$ times.
We collect results after each pass and plot in Figure~\ref{fig:res_restream}.
The first restream generates a considerable quality jump, with an improvement over the baseline of $24.6\%$ on average.
Each following pass has a positive impact on solution quality, which converges to be a $40.9\%$-improvement on average over the baseline after the last pass.
On the other hand, the running time has a roughly linear increase for each pass over the graph.
In practice, this adds another degree of freedom to configure \AlgName{HeiStream} for the needs of real systems.

\subsection{Comparison against State-of-the-Art}
\label{subsec:Scenarios}

\begin{table*}[b]
	\centering
	\footnotesize
	\vspace*{-.5cm}
	\caption{Edge-cut results against competitors for $k=32$. Internal memory algorithms are are on the 2 left columns and streaming algorithms are on the 5 right columns. We refer to setups of \AlgName{HeiStream} with specific buffer sizes as \AlgName{HS}($X$k), in which each buffer contains $X\times 1024$ nodes. \AlgName{HS}(Int.) uses a buffer size of $n$. We bold the best result for each graph for internal memory approaches and streaming approaches. The results for \AlgName{Multilevel~LDG} for the five bottom graphs are missing, as the graphs where not part of their benchmark set. Lower is better. }
	\vspace*{-.25cm}
	\begin{tabular}{|l|rr|rrrrr|}
		\hline
		
		\multicolumn{1}{|c|}{}                        & \multicolumn{7}{c|}{Cut Edges (\%)}                                 \\ \cline{2-8} 
		\multicolumn{1}{|c|}{\multirow{-2}{*}{ Graph}} & \AlgName{HS}(Int.) & \AlgName{MLDG}(Int.)  & \AlgName{HS}(32k) & $2$-\AlgName{ReFennel} & \AlgName{Fennel} & \AlgName{LDG}                       & \AlgName{Hashing} 
		\\

		\hline\hline

		Dubcova1               & \textbf{13.68}        & 14.26           & \textbf{13.68}    &         29.19    & 33.99  &         33.96  & 95.62   \\ 
		hcircuit               & \textbf{2.73}         & 17.75           & \textbf{2.53}     &         21.86    & 28.97  &         28.97  & 90.75   \\ 
		coAuthorsDBLP          & \textbf{15.99}        & 24.82           & \textbf{17.80}    &         24.28    & 27.12  &         27.12  & 94.80   \\ 
		Web-NotreDame          & \textbf{5.85}         & 11.01           & \textbf{9.20}     &         12.58    & 19.52  &         19.56  & 95.97   \\
		Dblp-2010              & \textbf{11.31}        & 18.52           & \textbf{13.42}    &         22.93    & 28.82  &         28.80  & 92.49   \\ 
		ML\_Laplace            & \textbf{7.93}         & 13.44           &         8.36      &         7.82     & 7.92   & \textbf{5.77}  & 96.37   \\ 
		coPapersCiteseer       & \textbf{8.23}         & 11.22           &         10.29     & \textbf{9.63}    & 12.88  &         12.27  & 96.52   \\ 
		coPapersDBLP           & \textbf{14.51}        & 19.29           &         18.33     & \textbf{16.47}   & 20.65  &         20.22  & 96.39   \\ 
		Amazon-2008            & \textbf{10.09}        & 19.01           & \textbf{15.56}    &         28.92    & 37.07  &         37.07  & 94.68   \\
		eu-2005                & \textbf{11.14}        & 14.57           & \textbf{18.64}    &         25.53    & 35.88  &         31.96  & 96.44   \\
		Web-Google             & \textbf{1.62}         & 9.66            & \textbf{9.48}     &         18.04    & 30.64  &         30.64  & 96.87   \\
		ca-hollywood-2009      &         35.34         & \textbf{32.51}  &         42.36     & \textbf{41.34}   & 44.54  &         45.25  & 96.62   \\
		Flan\_1565             & \textbf{7.69}         & 9.36            &         8.12      &         10.26    & 10.59  & \textbf{6.70}  & 96.61   \\ 
		Ljournal-2008          & \textbf{29.12}        & 34.76           & \textbf{38.58}    &         43.23    & 51.43  &         51.36  & 96.07   \\ 
		HV15R                  &         14.09         & \textbf{11.48}  &         17.48     & \textbf{15.05}   & 16.39  &         15.49  & 96.84   \\ 
		Bump\_2911             & \textbf{8.66}         & 11.73           & \textbf{8.23}     &         8.61     & 8.65   &         8.30   & 96.19   \\ 
		FullChip               & \textbf{38.20}        & 48.16           & \textbf{45.71}    &         57.39    & 61.93  &         64.23  & 95.06   \\ 
		patents                & \textbf{15.57}         & 29.12           &         60.56     & \textbf{52.60}   & 70.98  &         70.98  & 96.88   \\ 
		Cit-Patents            & \textbf{15.75}        & 28.65           &         60.76     & \textbf{51.62}   & 72.16  &         72.16  & 96.88   \\ 
		Soc-LiveJournal1       & \textbf{29.72}        & 35.69           &         35.62     & \textbf{34.00}   & 39.03  &         45.62  & 96.66   \\ 
		circuit5M              &         40.02         & \textbf{34.60}  & \textbf{41.00}    &         75.45    & 78.42  &         78.47  & 96.87   \\

		\hline

		del21                  & 1.38                  & -               & \textbf{8.53}     &         33.52    & 40.21  &         40.21  & 93.39   \\ 
		rgg21                  & 1.53                  & -               & \textbf{1.52}     &         4.09     & 5.02   &         4.88   & 96.89   \\ 
		soc-orkut-dir          & 37.86                 & -               &         54.76     & \textbf{47.22}   & 55.85  &         60.27  & 96.85   \\
		italy-osm              & 0.13                  & -               & \textbf{1.34}     &         4.65     & 4.80   &         4.80   & 78.11   \\ 
		great-britain-osm      & 0.16                  & -               & \textbf{1.63}     &         7.18     & 7.34   &         7.34   & 79.94   \\ \hline
		
	\end{tabular}
	\label{tab:multiLDG}
	\vspace*{-.25cm}
\end{table*}


\begin{sidewaysfigure}
	\centering
	\textcolor{white}{\rule{0.75\textheight}{0.7\textheight}}
	\includegraphics[width=\textheight]{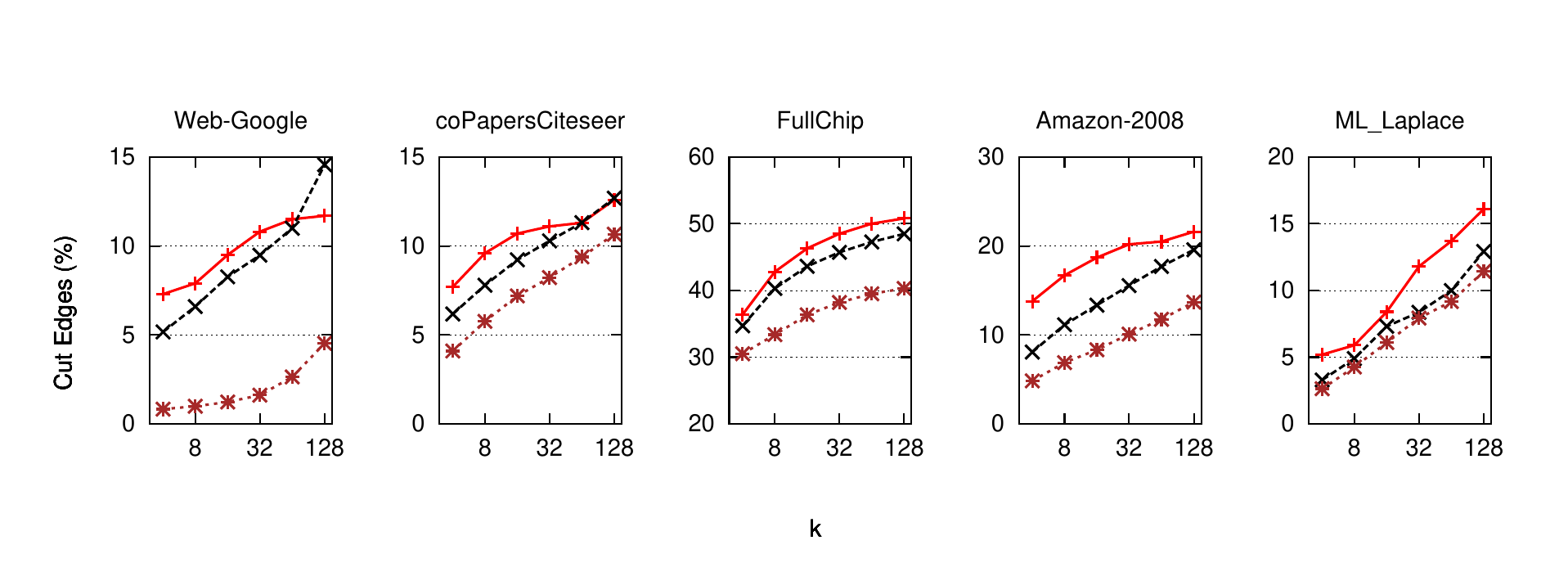}
	\caption{Comparison of \AlgName{HeiStream} against \AlgName{Multilevel~LDG} for buffer containing the whole graph.}
	\label{fig:multiLDG}
\end{sidewaysfigure}

\begin{figure*}[htb]
	\vspace*{-.5cm}
	\captionsetup[subfigure]{justification=centering}
	\centering
	\begin{subfigure}{\scaleFactor\textwidth}
		\centering
		\includegraphics[width=\imgScaleFactor\textwidth]{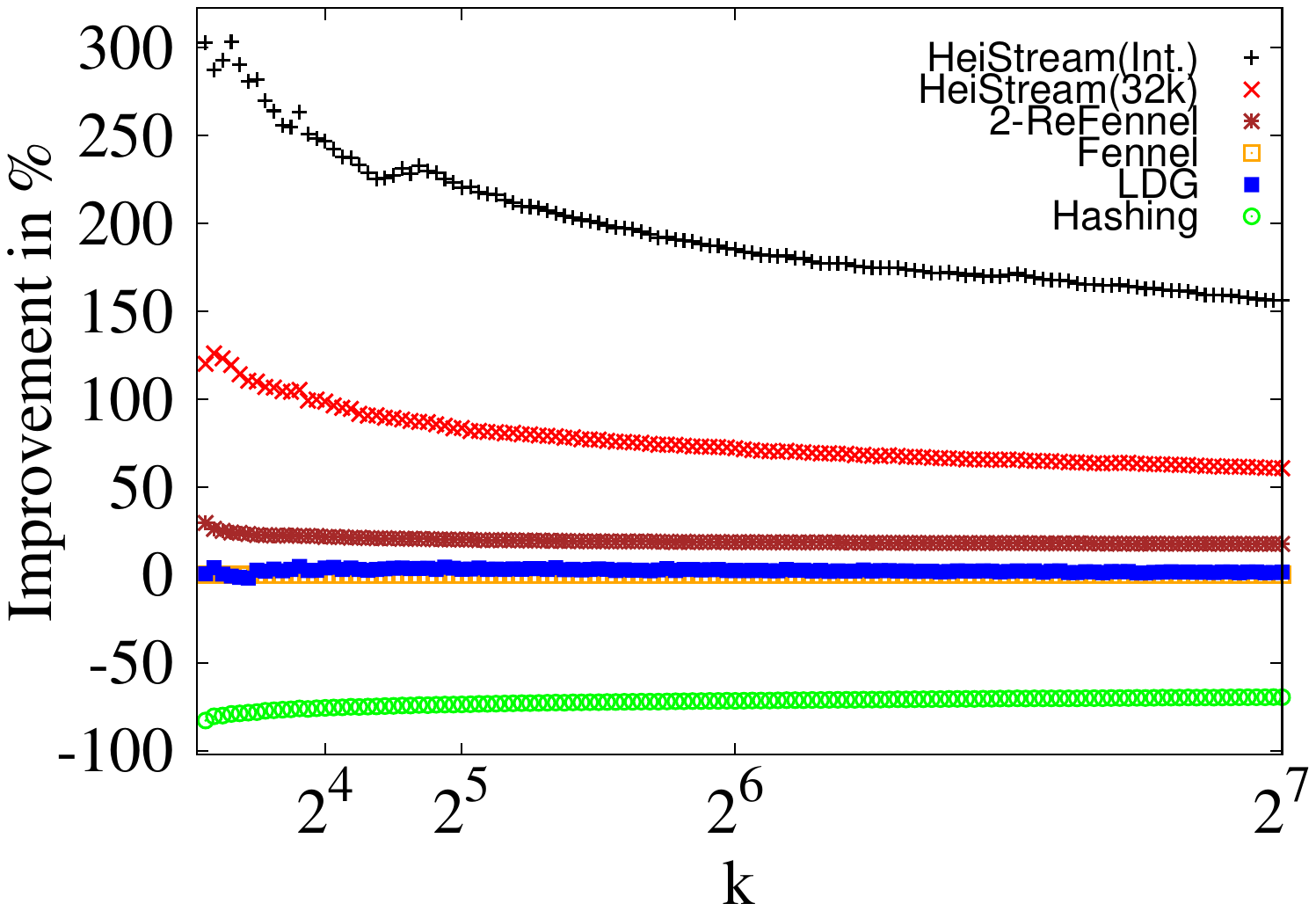}
		\vspace*{\capPosition}
		\caption{Quality improvement plot over \AlgName{Fennel}.}
		\label{fig:res_InitFennelOpt4}
	\end{subfigure}\hspace{5mm}
	\begin{subfigure}{\scaleFactor\textwidth}
		\centering
		\includegraphics[width=\imgScaleFactor\textwidth]{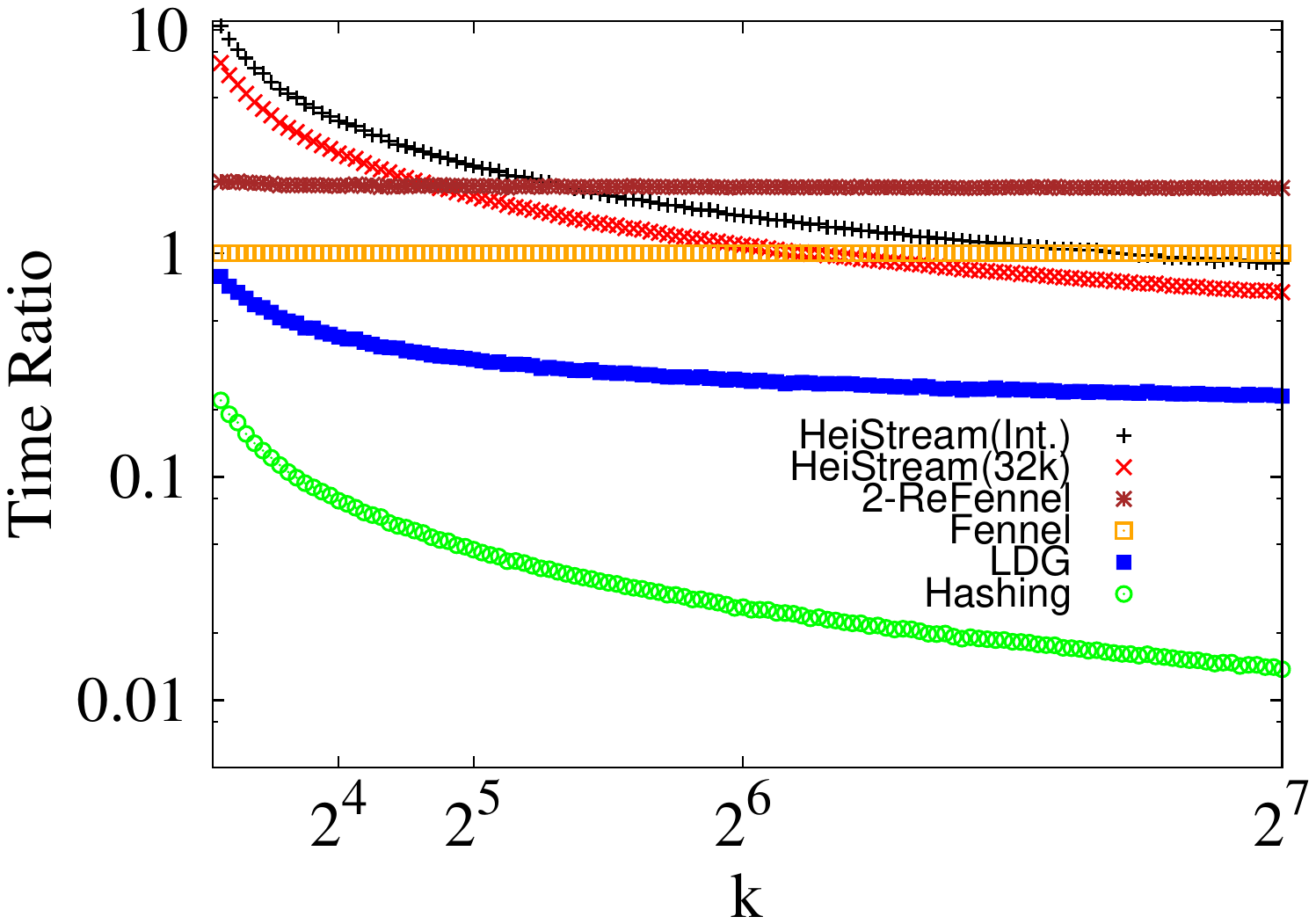}
		\vspace*{\capPosition}
		\caption{Relative running time plot over \AlgName{Fennel}.}
		\label{fig:tim_InitFennelOpt4}
	\end{subfigure}
	\vspace*{-.5cm}
	\caption{Results for comparison against state-of-the-art one-pass (re)streaming algorithms. Higher is better for quality improvement plots. Lower is better for running time.}
	\vspace*{-.5cm}
	\label{fig:one-pass1}
\end{figure*} 

\begin{figure*}[htb]
	
	\begin{subfigure}{\scaleFactor\textwidth}
		\centering
		\includegraphics[width=\imgScaleFactor\textwidth]{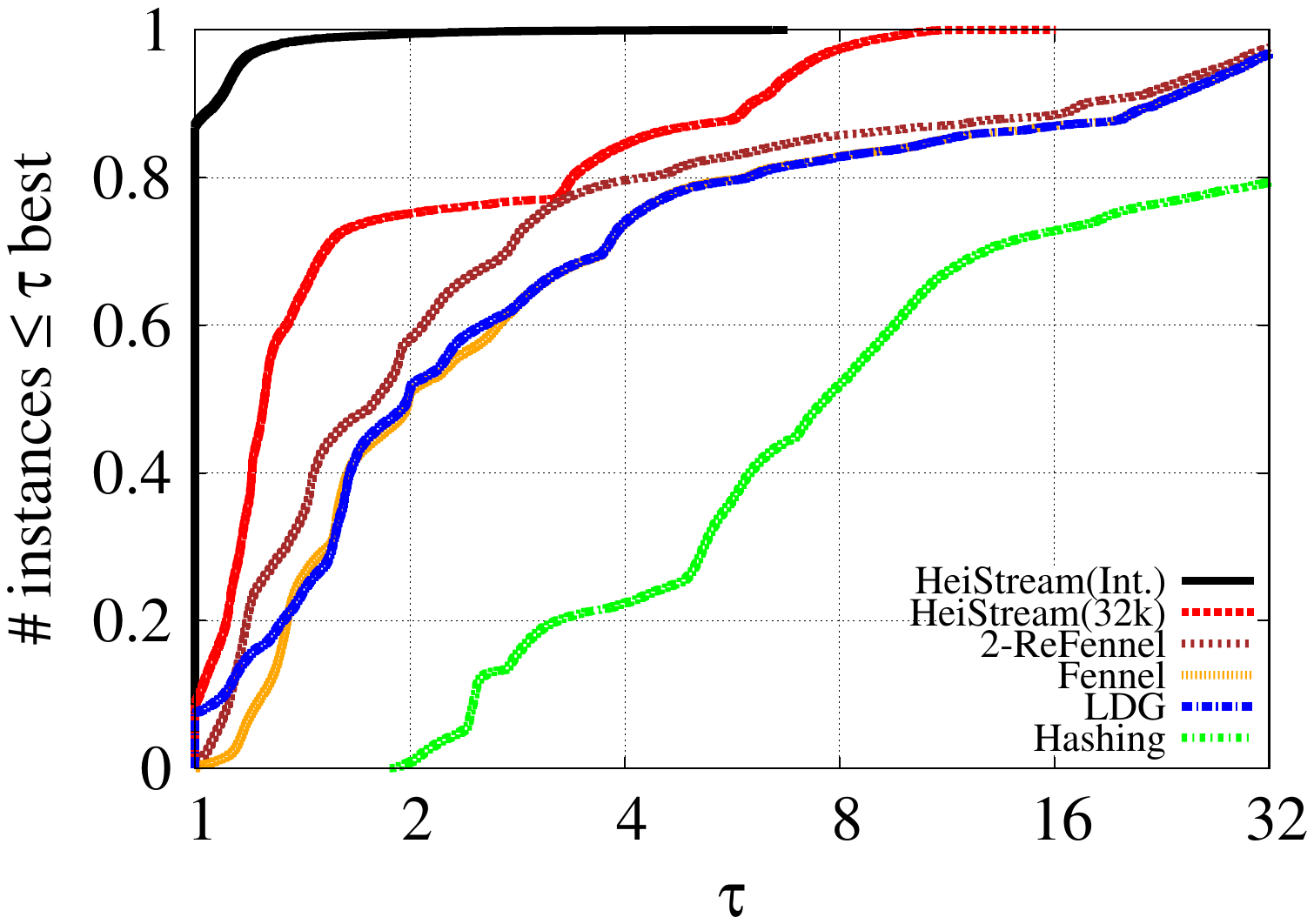}
		\vspace*{\capPosition}
		\caption{Quality performance profile.}
		\label{fig:pp_res_InitFennelOpt4}
	\end{subfigure}\hspace{5mm}
	\begin{subfigure}{\scaleFactor\textwidth}
		\centering
		\includegraphics[width=\imgScaleFactor\textwidth]{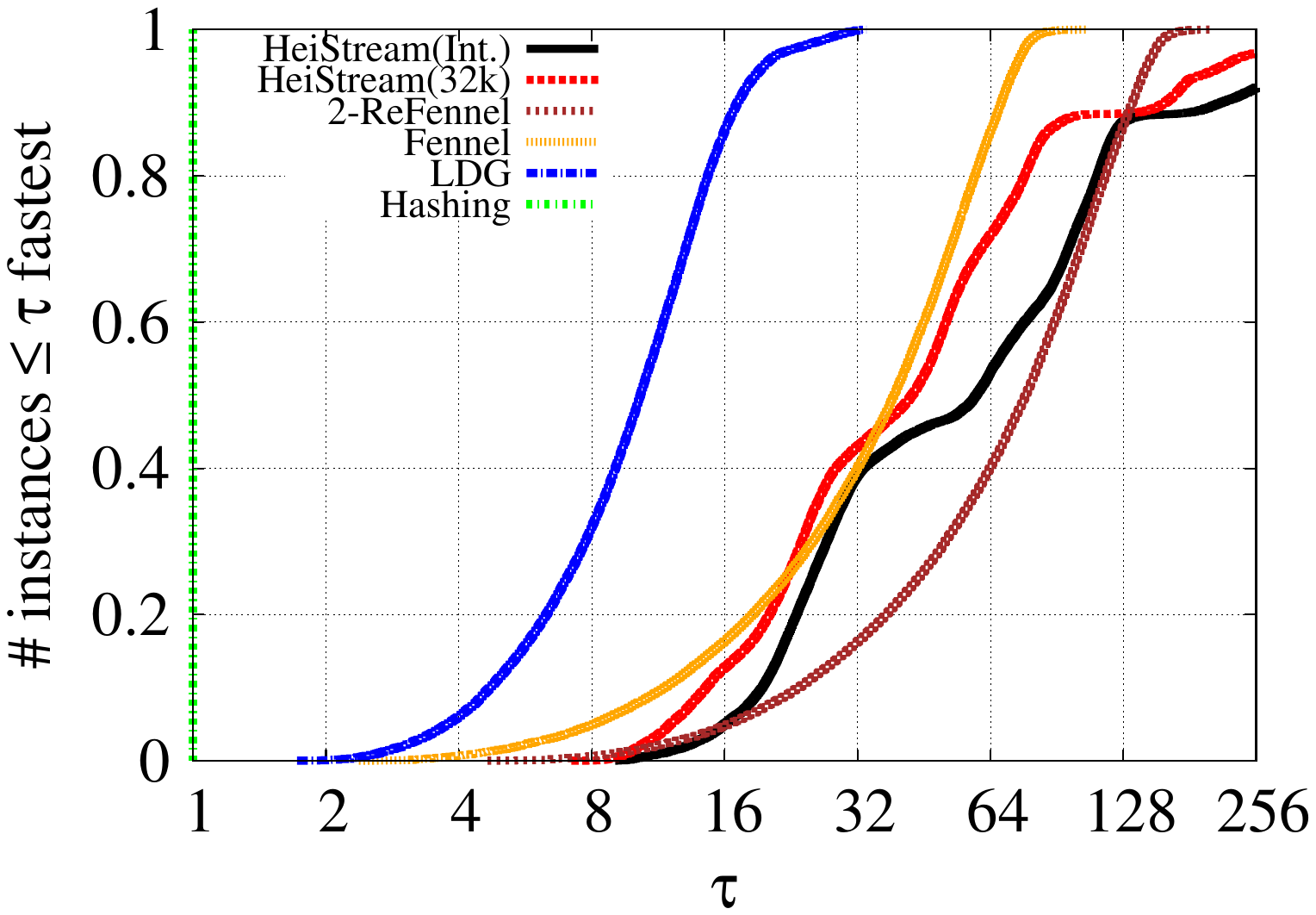}
		\vspace*{\capPosition}
		\caption{Running time performance profile.}
		\label{fig:pp_tim_InitFennelOpt4}
	\end{subfigure}
	\caption{Results for comparison against state-of-the-art one-pass (re)streaming algorithms using performance profiles. Higher is better for quality improvement plots. }
	\vspace*{-.5cm}
	\label{fig:one-pass2}
\end{figure*}

In this section, we show experiments in which we compare \AlgName{HeiStream} against the current state-of-the-art algorithms.
Unless mentioned otherwise, these experiments involve all the graphs from the Test Set group in Table~\ref{tab:graphs} and focus on two particular configurations of \AlgName{HeiStream}, which we refer to as \AlgName{HeiStream}(32k) and \AlgName{HeiStream}(Int.).
The first configuration is based on batches of size \numprint{32768}, while the second one has enough batch capacity to operate as an internal memory algorithm -- both configurations perform a single pass over the input using the extended~model.
Internal memory algorithms such as \AlgName{Metis}~\cite{karypis1998fast} and \AlgName{KaHIP}~\cite{kaffpa} are beyond the scope of this work since it is common knowledge that internal memory algorithms are better than streaming algorithms regarding partition quality if the instances fits in the memory of a machine.
For the sake of reproducibility, we ran \AlgName{Metis} and the fast social version of \AlgName{KaHIP} over our Test Set group of instances for $k \in \{2,4,8,16,32,64,128\}$.
On average, \AlgName{Fennel} cuts a factor $7.5$ more edges than \AlgName{Metis} and the fast social version of \AlgName{KaHIP}, while \AlgName{HeiStream}(Int.) cuts a factor $2.2$ more edges than both.
Furthermore, \AlgName{Fennel} is respectively a factor $2.2$ and a factor $6.0$ faster than \AlgName{Metis} and the fast social version of \AlgName{KaHIP}, while \AlgName{HeiStream}(Int.) is $56.3\%$ slower than \AlgName{Metis} and $72.9\%$ faster than the fast social version~of~\AlgName{KaHIP}.

\textbf{Results.}
We now present a detailed comparison of \AlgName{HeiStream} (HS) against the state-of-the-art.
In the results, we refer to the internal memory version of \AlgName{Multilevel~LDG} as \AlgName{MLDG}(Int.).
Moreover, we refer to the restreaming version of \AlgName{Fennel} that passes over the graph $2$ times as $2$-\AlgName{ReFennel}. 
First, we focus on $k=32$ and later on choose a much wider range for the number of blocks.
Table~\ref{tab:multiLDG} shows the percentage of edges cut in the partitions generated by each algorithm for the graphs in the Test Set for $k=32$.
\AlgName{HeiStream}(Int.) and \AlgName{HeiStream}(32k) outperform all the other competitors for the majority of instances.
First, both outperform \AlgName{Hashing} for all graphs and \AlgName{LDG} for $23$ out of the $26$ graphs.
Next, both outperform \AlgName{Fennel} in $24$ instances.
The algorithm $2$-\AlgName{ReFennel} is outperformed by \AlgName{HeiStream}(Int.) and \AlgName{HeiStream}(32k) in $24$ and $17$ instances, respectively.
Considering only the 21 instances for which there are results reported for \AlgName{Multi.LDG}(Int.) in literature, the algorithms \AlgName{HeiStream}(Int.) and \AlgName{HeiStream}(32k) compute better partitions for $18$~and~$14$ instances respectively.

For a closer comparison against \AlgName{Multi.LDG}(Int.), we present Figure~\ref{fig:multiLDG}.
We plot edge-cut values for \AlgName{Multi.LDG}(Int.) based on results graphically reported in~\cite{jafari2021fast}.
In this figure, we show of \AlgName{HeiStream}(Int.) and \AlgName{HeiStream}(32k) for $5$ particular graphs with $k=4,8,16,32,64,128$.
For all these instances, \AlgName{HeiStream}(Int.) outperforms \AlgName{Multi.LDG}(Int.) by a considerable margin.
Observe that \AlgName{HeiStream}(32k) outperforms the \emph{internal memory} version of \AlgName{Multilevel~LDG} for the majority of instances.
We omit additional comparisons against buffered versions of \AlgName{Multilevel~LDG}, since they provide lower quality than the internal memory version.
We ran wider experiments over our whole Test Set for $127$ different values of $k$.
Figure~\ref{fig:one-pass1} shows a quality improvement plot over \AlgName{Fennel} and a relative running time plot.
Figure~\ref{fig:one-pass2} shows performance profiles for solution quality and running~time.
Observe that \AlgName{HeiStream}(Int.) produces solutions with highest quality overall.
In particular, it produces partitions with smallest edge-cut for almost $86.9\%$ of the instances and improves solution quality over \AlgName{Fennel} $195.0\%$ on average. 
We now provide some results in which we exclude \AlgName{HeiStream}(Int.), since it has access to the whole graph.
The best algorithm is \AlgName{HeiStream}(32k), which produces the best solution quality for $63.3\%$ of the instances and improves solution quality over \AlgName{Fennel}  $75.9\%$ on average.
It is followed by $2$-\AlgName{ReFennel}, which is the best tested algorithm from the previous state-of-the-art. 
In particular, it computes the best partition for $26.8\%$ of the instances and improves on average $19.2\%$ over \AlgName{Fennel}.
\AlgName{LDG} and \AlgName{Fennel} come next.
Particularly, \AlgName{LDG} finds the best partition for $9.8\%$ of the instances and improves on average $2.4\%$ over \AlgName{Fennel}. 
\AlgName{Fennel} does not find the best partition for any instance.
Finally, \AlgName{Hashing} produces the worst solutions, with $72.5\%$ worse quality than \AlgName{Fennel} on average.
Regarding running time, \AlgName{Hashing} is the fastest one for all instances, which is expected since it is the only one with time complexity $O(n)$.
The second fastest one is \AlgName{LDG}, whose running time is a factor $9.6$ higher than \AlgName{Hashing} on average.
\AlgName{Fennel}, \AlgName{HeiStream}(32k) and \AlgName{HeiStream}(Int.) come next with factors $32.4$, $41.4$ and $56.2$ slower than \AlgName{Hashing} on average, respectively.
\AlgName{ReFennel} is the slowest algorithm of the test, being a factor $64.61$ slower than \AlgName{Hashing} on average.
In a direct comparison, \AlgName{HeiStream}(32k) and \AlgName{HeiStream}(Int.) are respectively $27.7\%$ and $73.2\%$ slower than \AlgName{Fennel} on average.
Note that both configurations of \AlgName{HeiStream} are faster than \AlgName{Fennel} for larger values of~$k$, which is consistent with the fact the running time of \AlgName{HeiStream} is $O(n+m)$ while the running time of \AlgName{Fennel}~is~$O(nk+m)$. 

\begin{figure}[t]
	\centering
	\includegraphics[width=0.96\textwidth]{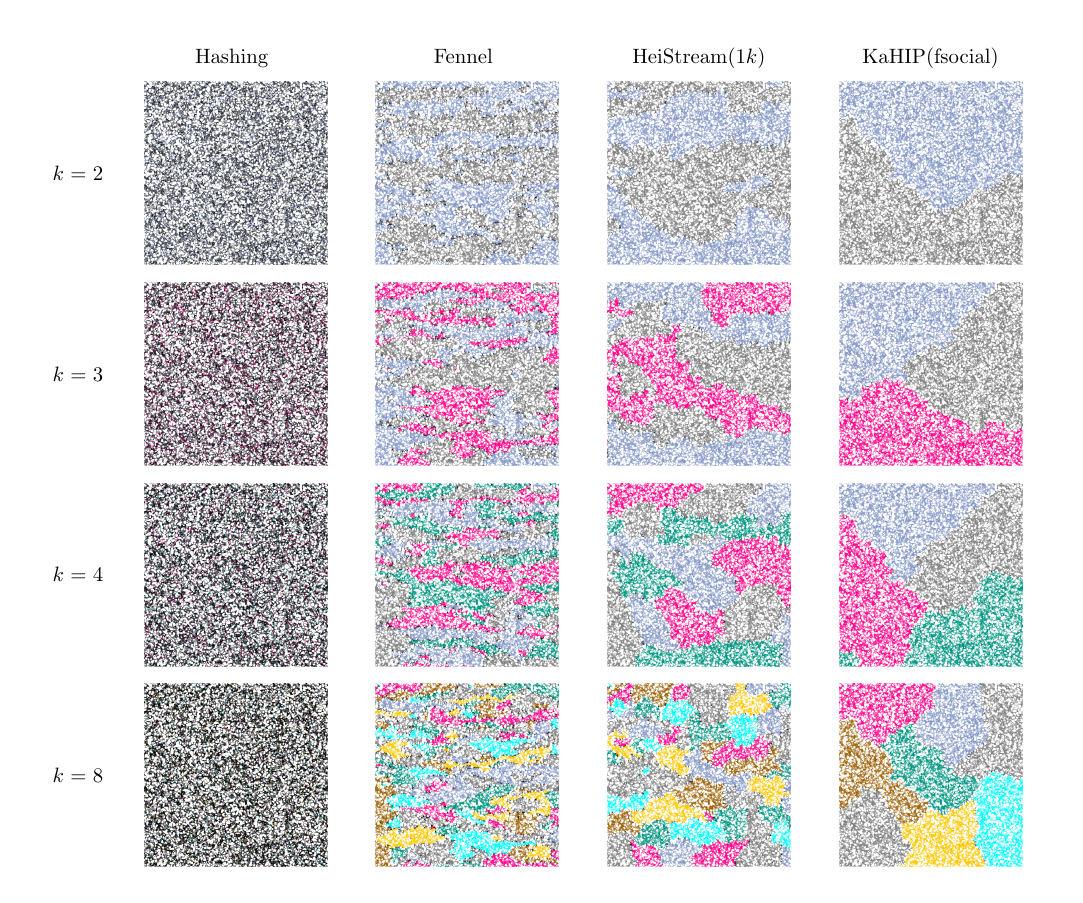}
	\caption{Visualization of partitions generated by \AlgName{Hashing}, \AlgName{Fennel}, \AlgName{HeiStream($1k$)} and the fast social version of \AlgName{KaHIP} for the graph rgg15, which has \numprint{32768} nodes and \numprint{160240} edges.}
	\label{fig:visualization}
\end{figure}

\textbf{Visualization.}
As shown, the edge-cut of partitions produced by \AlgName{HeiStream} is on average lower than the edge-cut of partitions produced by its competitor streaming algorithms.
We shortly look at some visualizations in order to concretely understand why this happens.
In Figure~\ref{fig:visualization}, we show a visual comparison of some partition layouts generated by \AlgName{Hashing}, \AlgName{Fennel}, \AlgName{HeiStream} and the fast social version of \AlgName{KaHIP} for the graph rgg15.
Since this graph has only \numprint{32768} nodes, we use a buffer size of \numprint{1024} nodes for \AlgName{HeiStream} in order to partition the graph over multiple successive batches.
There is a leap of partitioning quality from \AlgName{Hashing} to \AlgName{Fennel}, \ie well-delimited clusters associated to a same block can be identified in the partitions generated by \AlgName{Fennel} but not in partitions generated by \AlgName{Hashing}.
Similarly there is a leap of partitioning quality from \AlgName{Fennel} to \AlgName{KaHIP}, \ie a block generated by \AlgName{Fennel} consists of multiple small clusters that are not mutually connected while a block generated by \AlgName{KaHIP} usually consists of a single connected cluster.
Note that the partitions produced by \AlgName{HeiStream} have intermediary characteristics between the partitions generated by \AlgName{Fennel} and \AlgName{KaHIP}.
More specifically, a block generated by \AlgName{HeiStream} consists of fewer and larger clusters than a block generated by \AlgName{Fennel} but not as few and as large clusters as those generated by \AlgName{KaHIP}.
This behavior is a direct consequence of the more or less global view provided by the distinct computational models used by these three algorithms.

\subsection{Huge Graphs}
\label{sec:huge_graphs}

\begin{table*}[p]
	\centering
	\footnotesize
	\caption{Results of experiments with the huge graphs from Table \ref{tab:graphs}. For each instance, we present the percentage of cut edges (CE) and the running time in seconds (RT). We bold the best result for each instance. We refer to setups of \AlgName{HeiStream} with specific buffer sizes as \AlgName{HeiStream}($X$k), in which each buffer contains $X\times 1024$ nodes.
	}
	\hspace*{-.25cm}	\begin{tabular}{|lr|rrr|rr|rr|rr|}
		\hline
		\multicolumn{1}{|c}{\multirow{2}{*}{Graph}} & \multirow{2}{*}{k} & \multicolumn{3}{c|}{\AlgName{HeiStream}(Xk)}                                               & \multicolumn{2}{c|}{\AlgName{Fennel}}                               & \multicolumn{2}{c|}{\AlgName{LDG}}                                  & \multicolumn{2}{c|}{\AlgName{Hashing}}                              \\ \cline{3-11} 
		\multicolumn{1}{|c}{}                       &                    & \multicolumn{1}{c}{X} & \multicolumn{1}{c}{CE(\%)} & \multicolumn{1}{c|}{RT(s)} & \multicolumn{1}{c}{CE(\%)} & \multicolumn{1}{c|}{RT(s)} & \multicolumn{1}{c}{CE(\%)} & \multicolumn{1}{c|}{RT(s)} & \multicolumn{1}{c}{CE(\%)} & \multicolumn{1}{c|}{RT(s)} \\ 
		
		\hline\hline

		\multirow{6}{*}{uk-2005}                   
		& 8                  & 1024                  & \textbf{4.03}              & 290.23                      & 19.93                      & 37.19                       & 19.97                      & 19.36                       & 73.70                      & 3.28                         \\
		& 16                 & 1024                  & \textbf{6.01}              & 300.04                      & 22.76                      & 58.05                      & 22.72                      & 22.58                       & 78.86                      & 3.38                         \\
		& 32                 & 1024                  & \textbf{7.65}              & 310.72                      & 25.24                      & 98.78                       & 25.19                      & 29.19                       & 81.39                      & 3.30                         \\
		& 64                 & 1024                  & \textbf{8.99}              & 322.14                      & 26.88                      & 183.15                      & 26.81                      & 42.90                       & 82.60                      & 3.28                         \\
		& 128                & 1024                  & \textbf{9.94}              & 346.73                      & 27.89                      & 342.74                      & 27.76                      & 61.87                       & 83.18                      & 3.27                         \\	
		& 256                & 1024                  & \textbf{10.68}              & 386.64                      & 28.78          &  666.22           &            28.65           &   109.20                      &  83.46                     &   3.31                       \\ \hline
		
		\multirow{6}{*}{twitter7}                  
		& 8                  & 512                   & \textbf{41.64}             & 1727.13                      &      45.18                 & 184.17                       & 56.11                      & 180.85                       & 71.66                      & 3.46                         \\
		& 16                 & 512                   & \textbf{47.04}             & 1774.92                      & 53.49                      & 213.17                       & 61.73                      & 186.27                       & 76.78                      & 3.57                        \\
		& 32                 & 512                   & \textbf{52.59}             & 1884.16                      & 58.15                      & 244.18                       & 66.84                      & 184.90                       & 79.34                      & 3.49                        \\
		& 64                 & 512                   & \textbf{57.53}             & 1988.11                      & 62.95                      & 330.46                      & 68.68                      & 197.86                       & 80.62                      & 3.50                        \\
		& 128                & 512                   & \textbf{61.87}             & 2113.34                      & 66.68                      & 504.53                      & 69.94                      & 219.65                       & 81.26                      & 3.93                         \\
		& 256                & 512                   & \textbf{65.47}             & 2357.92                      & 78.32                      & 846.20                     & 71.26                      & 280.99                    & 81.57                     & 3.51                         \\ \hline

		\multirow{6}{*}{sk-2005}                   
		& 8                  & 1024                  & \textbf{3.23}              & 634.79                      & 21.95                      & 55.39                       & 21.13                      & 30.98                       & 81.50                      & 4.17                        \\
		& 16                 & 1024                  & \textbf{4.11}              & 648.48                      & 26.36                      & 82.41                       & 25.33                      & 35.44                       & 87.26                      & 4.20                        \\
		& 32                 & 1024                  & \textbf{5.32}              & 667.84                      & 29.59                      & 137.42                      & 27.97                      & 43.76                       & 90.11                      & 4.23                        \\
		& 64                 & 1024                  & \textbf{7.55}              & 695.60                      & 32.52                      & 238.54                      & 30.18                      & 59.19                       & 91.50                      & 4.21                        \\
		& 128                & 1024                  & \textbf{8.95}              & 733.05                      & 35.87                      & 449.19                      & 32.44                      & 91.36                       & 92.19                      & 4.20                        \\ 
		& 256                & 1024                  & \textbf{12.02}              & 798.73                     & 40.06                     & 857.76                    &  35.69                    &        150.64               & 92.55                     & 4.26                     \\ \hline
		
		\multirow{6}{*}{soc-friendster}           
		& 8                  & 1024                  & \textbf{27.36}             & 4099.35                     & 30.57                      & 405.68                       & 45.60                      & 381.78                       & 87.53                      & 5.45                        \\
		& 16                 & 1024                  & \textbf{34.50}             & 4202.04                     & 45.74                      & 440.11                       & 58.98                      & 361.54                       & 93.77                      & 5.62                        \\
		& 32                 & 1024                  & \textbf{39.52}             & 4345.96                     & 54.87                      & 503.90                      & 61.00                      & 408.56                       & 96.89                      & 5.49                        \\
		& 64                 & 1024                  & \textbf{46.35}             & 4546.98                     & 59.27                      & 649.34                      & 64.02                      & 422.87                       & 98.45                      & 5.44                        \\
		& 128                & 1024                  & \textbf{52.41}             & 4796.56                     & 60.82                      & 888.14                      & 68.17                      & 475.16                       & 99.22                      & 5.72                        \\ 
		& 256                & 1024                  & \textbf{57.79}             & 5323.08                     & 64.25                      & 1426.16            & 71.90                & 523.66               &        99.61               &  5.53                       \\ \hline

		\multirow{6}{*}{er-fact1.5s26}           
		& 8                  & 1024                  & \textbf{73.27}             & 2216.99                      & 73.44                      & 259.98                       & 73.44                      & 208.81                       & 87.50                      & 5.57                        \\
		& 16                 & 1024                  & \textbf{80.18}             & 2292.12                      & 80.40                      & 288.90                       & 80.40                      & 226.01                       & 93.75                      & 5.57                        \\
		& 32                 & 1024                  & \textbf{84.36}             & 2400.35                     & 84.63                      & 357.21                      & 84.63                      & 255.60                       & 96.87                      & 5.60                        \\
		& 64                 & 1024                  & \textbf{86.99}             & 2534.09                     & 87.31                      & 506.23                      & 87.31                      & 270.58                       & 98.44                      & 5.58                        \\
		& 128                & 1024                  & \textbf{88.72}             & 2725.81                     & 89.10                      & 769.61                      & 89.10                      & 407.59                       & 99.22                      & 5.57                        \\ 
		& 256                & 1024                  & \textbf{89.99}             & 2913.95                     & 90.45                      & 1329.57                 & 90.45                      &      408.35                 & 99.61                      & 5.65                        \\ \hline

		\multirow{6}{*}{RHG1}                     
		& 8                  & 1024                  & \textbf{0.04}              & 380.04                      & 2.02                       & 86.95                       & 2.02                       & 44.42                       & 91.91                      & 8.37                        \\
		& 16                 & 1024                  & \textbf{0.06}              & 391.63                      & 2.12                       & 143.47                      & 2.12                       & 52.11                       & 97.39                      & 8.57                        \\
		& 32                 & 1024                  & \textbf{0.09}              & 406.56                      & 2.16                       & 252.15                      & 2.16                       & 65.65                       & 99.19                      & 8.38                        \\
		& 64                 & 1024                  & \textbf{0.15}              & 435.71                      & 2.17                       & 450.73                      & 2.17                       & 95.22                       & 99.74                      & 8.33                        \\
		& 128                & 1024                  & \textbf{0.22}              & 482.06                      & 2.18                       & 877.69                      & 2.18                       & 147.22                      & 99.90                      & 8.31                        \\
		& 256                & 1024                  & \textbf{0.34}              & 569.77                     &   2.19                    &   1708.33                 &   2.18                     &   273.76                   &   99.95                    &   8.45                   \\ \hline

		\multirow{6}{*}{RHG2}                      
		& 8                  & 1024                  & 0.09                       & 621.56                      & 0.05                       & 103.83                       & \textbf{0.04}              & 56.23                       & 89.71                      & 8.31                        \\
		& 16                 & 1024                  & 0.13                       & 632.61                      & 0.07                       & 153.29                      & \textbf{0.04}              &    60.14                         & 96.15                      & 8.57                        \\
		& 32                 & 1024                  & 0.19                       & 648.68                      & 0.12                       & 262.91                      & \textbf{0.05}              & 77.34                       & 98.73                      & 9.85                        \\
		& 64                 & 1024                  & 0.29                       & 674.36                      & 0.18                       & 468.04                      & \textbf{0.05}              & 108.39                       & 99.57                      & 8.32                        \\
		& 128                & 1024                  & 0.44                       & 727.66                      & 0.28                       & 872.68                      & \textbf{0.07}              & 157.75                      & 99.85                      & 8.31                        \\
		& 256                & 1024                  & 0.68                       & 816.60                      & 0.44                       & 1686.18                   & \textbf{0.09}              & 278.51                     & 99.92                      & 8.47                      \\ \hline

		\multirow{6}{*}{uk-2007-05}                
		& 8                  & 1024                  & \textbf{0.54}              & 1024.26                      & 25.23                      & 107.46                       & 25.21                      & 58.19                       & 87.91                      & 8.80                        \\
		& 16                 & 1024                  & \textbf{0.60}              & 1045.36                      & 28.02                      & 166.06                      & 28.19                      & 70.72                       & 94.08                      & 8.79                        \\
		& 32                 & 1024                  & \textbf{0.70}              & 1058.73                      & 29.40                      & 278.38                       & 29.32                      & 85.42                       & 97.12                      & 8.99                        \\
		& 64                 & 1024                  & \textbf{0.92}              & 1099.64                      & 29.94                      & 517.20                      & 29.85                      & 115.18                       & 98.61                      & 9.32                        \\
		& 128                & 1024                  & \textbf{1.31}              & 1163.45                      & 30.73                      & 935.98                      & 30.18                      & 175.01                      & 99.33                      & 8.79                        \\
		& 256                & 1024                  & \textbf{1.95}              & 1280.48                      & 31.65                      & 1808.52                    & 30.70                      & 324.71                    & 99.68                      & 8.92                      \\ \hline

	\end{tabular}
	\label{tab:hugeResults}
\end{table*}

We now switch to the main use case of streaming algorithms: computing high-quality partitions for huge graphs on small machines. 
The experiments in this section are based on the \emph{huge graphs} listed in Table~\ref{tab:graphs} and are run on the relatively small Machine~B.
Namely, we ran experiments for $k=\{8,16,32,64,128,256\}$ 
and we did not repeat each test multiple times with different seeds as in previous experiments. We also ran \AlgName{Metis} and \AlgName{KaHIP} on those graphs on this machine, but they fail on all instances since they require more memory than the machine has.
For all instances, \AlgName{HeiStream} performs a single pass over the input based on the extended model construction.
We refer to setups of \AlgName{HeiStream} with specific buffer sizes as \AlgName{HeiStream}($X$k), in which a buffer contains $X\times 1024$ nodes. 
In~Table~\ref{tab:hugeResults}, we report detailed per-instance results with large buffer sizes able to run on Machine~B.
We exclude from Table~\ref{tab:hugeResults} the IO~delay to load the input graph from the disk, since it depends on the disk and is roughly the same independently of the used partitioning algorithm.
For completeness, we report this delay (in seconds) for the huge graphs listed in Table~\ref{tab:graphs} following their respective order: 131.3, 203.2, 313.2, 294.0, 164.9, 186.1, 340.7, 551.5.

The results show that \AlgName{HeiStream} outperforms all the competitor algorithms regarding solution quality for most instances.
Notably, \AlgName{HeiStream} computes partitions with considerably lower edge-cut in comparison to the one-pass algorithms for 4 out of the tested graphs: uk-2005, sk-2005, uk-2007-05, and RHG1.
For the social networks soc-friendster and twitter7, \AlgName{HeiStream} is the best for all instances, 
but the improvement over \AlgName{Fennel} and \AlgName{LDG} is not so large as in the other instances.
One outlier can be seen on the network RHG2.
While \AlgName{HeiStream} produces fairly small edge-cut values, which are all below $0.7\%$, \AlgName{Fennel} does outperform both and \AlgName{LDG} improves solution quality even further on this instance.
Furthermore, note that the running time of \AlgName{Fennel} increases with increasing $k$ to the point in which it becomes higher than the running time of \AlgName{HeiStream} for 5 out of the 8 huge graphs tested.
\paragraph*{Memory Consumption.} We now shortly review the amount of memory needed by the streaming algorithms under consideration. First of all note that the memory of \AlgName{HeiStream} depends on the size of the buffer that is used. If the buffer only contains one node, then the memory requirements match those of \AlgName{Fennel} and \AlgName{LDG}. Here, we measure the memory consumption of \AlgName{HeiStream} for various buffer sizes and compare it to \AlgName{Fennel} and \AlgName{LDG}.  To do that, we measured the memory consumption of \AlgName{HeiStream}(1024k), \AlgName{HeiStream}(32k), \AlgName{Fennel} and \AlgName{LDG} on the three largest graphs (RHG1, RHG2, and uk-2007-05). On average, \AlgName{HeiStream}(1024k) consumes respectively $2.5GB$, $4.1GB$, and $9.9GB$  of memory to partition these graphs. while \AlgName{HeiStream}(32k) consumes $472MB$, $521MB$, and $3.7GB$, and \AlgName{Fennel} and \AlgName{LDG} use $399MB$, $401MB$, and $445MB$. Note that the increased amount of used memory of \AlgName{HeiStream}(1024k) is expected, since we use a fairly large buffer. Given the size of the graphs, we believe those required amounts of memory are more than reasonable.

\section{Conclusion}
\label{s:conclusion}

We proposed \AlgName{HeiStream}, a buffered streaming graph partitioning algorithm.
We combined the buffered streaming model with multilevel graph partitioning techniques and extended \AlgName{Fennel} to a multilevel algorithm.
Compared to the previous state of-the-art of streaming graph partitioning, \AlgName{HeiStream} computes significantly better solutions than known streaming algorithms while at the same time being faster in many cases.
Note that improved edge cuts directly improve the communication cost in applications such as online queries on distributed graph databases~\cite{pacaci2019experimental}. Hence, our results directly yield improvements in applications. 
An important property of \AlgName{HeiStream} is that its running time does not depend on the number of blocks, while the previous state-of-the-art streaming partitioning algorithms have running time almost proportional to this number of blocks. 

\begin{acks}
Partially supported by DFG grant SCHU 2567/1-2.
\end{acks}

\bibliographystyle{ACM-Reference-Format}
\bibliography{phdthesiscs}

%
%
%
%
%
%
%
%

\end{document}